\documentclass[twocolumn,10pt]{article}

\usepackage{stfloats}

\usepackage{graphicx}
\usepackage{balance}
\usepackage{authblk}
\usepackage{natbib}
\usepackage[dvipsnames]{xcolor}
\usepackage{textgreek}
\usepackage[utf8]{inputenc}
\usepackage[english]{babel}
\usepackage{aas_macros}

\usepackage{enumitem} 

\usepackage{hyperref}
\hypersetup{
    unicode,
    colorlinks=true,
    linkcolor=linkcolor,
    citecolor=linkcolor,
    filecolor=linkcolor,
    urlcolor=linkcolor,
}
\usepackage{color,colortbl}
\definecolor{linkcolor}{rgb}{0.0,0.3,0.5}
\usepackage{orcidlink}

\usepackage{amsmath}
\usepackage{amssymb}
\usepackage{bm}
\usepackage{xspace}

\usepackage[ruled,vlined,linesnumbered]{algorithm2e}
\SetKwInOut{Require}{Require}
\SetKwInOut{Ensure}{Ensure}
\SetKwFor{ForPar}{for}{in parallel}{end}
\SetKwComment{tcp}{$\triangleright$\ }{}
\usepackage{algpseudocode}

\usepackage{booktabs}
\usepackage{multirow}

\DeclareMathOperator{\sinc}{sinc}
\newcommand{\nufftcf}{\texttt{nufftcf}\xspace}
\newcommand{\pastas}{\texttt{pastas}\xspace}
\newcommand{\pyzdcf}{\texttt{pyZDCF}\xspace}

\newcommand{\inlineeqnum}[1]{%
	\refstepcounter{equation}%
	\label{#1}%
	\text{(\theequation)}%
}

\graphicspath{ {./figs/} }

\title{nufftcf: Fast Auto- and Cross-Correlation Function Estimation for
	Irregularly-Sampled Time Series via the Non-Uniform FFT}
\author{Jean-Eric Campagne\orcidlink{0000-0002-1590-6927}}
\affil{Universit\'e Paris-Saclay, CNRS/IN2P3, IJCLab, 91405 Orsay, France \\ \texttt{jean-eric.campagne@ijclab.in2p3.fr}}
\date{\today}
\begin{document}
\twocolumn[ \begin{@twocolumnfalse}
\maketitle
\begin{abstract}
Estimating the auto-correlation function (ACF) and the cross-correlation function (CCF)
of sampled series is a standard task across different physics fields, e.g., astronomy
and environmental fields. We aim to provide an ACF/CCF estimator primarily designed for
irregularly sampled series that is numerically consistent with established,
well-validated kernel-weighted definitions, while scaling as $O(nK)$ for $K$ requested
lags (i.e., linearly in the series length $n$ at fixed $K$) rather than $O(n^2)$.

\nufftcf was designed first to evaluate the Wiener--Khinchin theorem for irregularly
sampled data using the Non-Uniform Fast Fourier Transform (NUFFT), via the Flatiron
Institute's \texttt{FINUFFT} library, and using the Gaussian- and rectangle-kernel
estimators. An $O(n)$-per-lag two-pointer scan replaces the naive $O(n^2)$ computation
of the effective pair count that normalizes each lag bin. Thanks to the pair-counting
algorithm development, we have designed real-space estimators with the same kernels that
have the same complexity and can be used as an exact reference, also faster than the
NUFFT path in some cases. The library also provides dedicated classical-FFT estimators
for regularly sampled data. All the estimators share a single calling convention that
eases switching from one to another.

With synthetic time series, we validate \nufftcf for ACF against \pastas, a library
known in groundwater time series analysis, and for CCF against \pyzdcf, known in
astronomical time series analysis. Benchmarks confirm the expected asymptotic scalings
--- $O(nK)$ for \nufftcf (NUFFT and real-space paths), with $K$ the number of requested
lags, versus $O(n^2)$ for the \pastas slotting technique --- with \nufftcf already
advantageous at moderate series lengths thanks to its low (millisecond-scale) overhead.
On regularly sampled data, the dedicated FFT path further reduces both cost and overhead
relative to the other estimators.

Beyond this cross-validation exercise, we further demonstrate \nufftcf on an
astrophysical use-case: a simulated ground-based stellar light curve combining a
quasi-periodic rotation signal, correlated flicker noise, seasonal sampling gaps, and
heteroscedastic measurement errors. Notebooks and scripts have been developed to help
users replay the examples shown in this article as well as others that go beyond.
\end{abstract}
\noindent \textbf{Keywords:} Time series analysis, Correlation function, Non-uniform
Fast Fourier Transform
\vspace{1em}
\end{@twocolumnfalse} 
]
\section{Introduction}
\label{sec:introduction}
Many measurements of physical, environmental, and astrophysical systems take the form of
a scalar time series that is not regularly sampled in time. Ground-based photometric
monitoring of active galactic nuclei (AGN) is interrupted by weather, daylight, and
telescope scheduling; groundwater levels are logged during irregular field visits;
paleoclimate proxies are deposited at a rate that itself varies with the climate signal
being measured. A recurring analysis task for such data is to estimate the
autocorrelation function (ACF) of a single series, or the cross-correlation function
(CCF) between two independently and irregularly sampled series, at a set of time lags
$\tau$.

Applications range from measuring a characteristic variability or persistence timescale,
to detecting a time delay between two related signals. For instance in AGN reverberation
mapping, where the CCF between a driving continuum light curve and a reprocessed
emission-line light curve encodes the size of the emitting region
\citep{2024HEAD...2110218W,asna.20210090,2021AJ....162..206S,
2019MNRAS.485.4790S,2014AdSpR..54.1414K}, one uses the discrete correlation function
\citep{edelson1988}, and also for instance the python version (\pyzdcf) of the code by
\cite{alexander97}. A comparable need arises in hydro(geo)logical time series analysis:
the open-source Python package Pastas \citep{collenteur2019}, widely used for time
series modeling of groundwater levels, implements an ACF estimator for irregularly
sampled series based on \cite{Rehfeld2011} and \cite{edelson1988}.

Because the classical, textbook estimator of the ACF/CCF assumes a common, regular
sampling grid \citep[e.g.][]{astsa2026}, irregular sampling has historically been
handled with different technics \citep[see][]{Kreutzer2013,Rehfeld2011}: resampling the
data onto a regular grid by interpolation before applying the standard Fourier-based
estimator; binning observation pairs by their time-lag separation (``slotting'');
weighting observation pairs by a smooth kernel of the time-lag separation rather than
binning them; or bypassing the lag domain altogether and working with a Fourier-domain
estimator adapted to irregular sampling, such as the Lomb--Scargle periodogram
\citep{2018ApJS..236...16V,1982ApJ...263..835S,1976Ap&SS..39..447L}.

Each of these approaches has been implemented, refined, and benchmarked by different
research communities---astronomy, geoscientific context among them---largely
independently of one another (Section~\ref{sec:related_methodology}). What these
approaches share is a computational cost that scales at best linearly per output lag
and, in the pairwise binning/weighting case, quadratically overall in the number of
observations $n$. This is unproblematic for the short, sparse times series where these
methods were originally designed for (typically $n \sim 10^2$--$10^3$), but certainly
would become a practical bottleneck for the longer, denser irregular time series
increasingly produced by continuous environmental monitoring, high-cadence time-domain
surveys, or long instrumental records ($n \sim 10^4$--$10^6$), and for workflows that
require many repeated evaluations (parameter sweeps, bootstrap resampling, batch
processing of large numbers of light curves or well records).

In this paper we present \nufftcf, an open-source Python package that computes ACF/CCF
estimates by evaluating the Wiener--Khinchin theorem with the Non-Uniform Fast Fourier
Transform (NUFFT) and classical-FFT for irregularly- and regularly-sampled time series
respectively. \nufftcf additionally provides direct, ``real-space'' implementations of
the same estimators as a reference and still valuable when only a few lags are
requested. All estimator families share a single calling convention, so that a user can
trade accuracy against speed without altering downstream analysis code.

The development of \nufftcf yields the following contributions:
\begin{enumerate}
    \item We demonstrate that the Gaussian- and rectangle-kernel ACF/CCF
    estimators---previously established on statistical grounds \citep{Rehfeld2011}---can
    be reformulated as a pair of Non-Uniform Fast Fourier Transform (NUFFT) evaluations.
    This reformulation ensures numerical consistency with existing definitions while
    replacing the pairwise sum of the numerator by a spectral evaluation costing
    $O(n\log n+K)$; the overall cost, dominated by the normalization of the next item,
    is $O(nK)$.
    \item We introduce an $O(n)$-per-lag analytical algorithm, based on a two-pointer
    scan, to compute the effective pair count for normalizing each lag bin. This
    replaces the naive $O(n^2)$ all-pairs approach, significantly improving efficiency,
    at the price of an overall $O(nK)$ cost for $K$ lags.
    \item We validate \nufftcf through comparisons with \pastas \citep{collenteur2019}
    and \pyzdcf \citep{jankov2022pyzdcf} on synthetic ACF and CCF problems, and
    benchmark their timings, including the number of lags $K^*$ above which the NUFFT
    estimators outperform the real-space ones.
\end{enumerate}

The paper is organized as follows. Section~\ref{sec:related_methodology} reviews prior
approaches to ACF/CCF estimation for irregularly-sampled data, situates \nufftcf
relative to them, and details the methodology, from the regularly-sampled case through
the four established families of irregular-sampling estimators to the NUFFT
reformulation used by \nufftcf. Section~\ref{sec:comparison} compares \nufftcf to
\pastas and \pyzdcf on ACF and CCF use cases. Section~\ref{sec:astro_example}
illustrates \nufftcf on a simulated astrophysical light curve closer to observational
practice. Section~\ref{sec:benchmarks} presents timing benchmarks.
Section~\ref{sec:discussion} discusses limitations and guidance on estimator choice, and
Section~\ref{sec:conclusion} concludes. The Appendix~\ref{app:algorithms} provides
additional details on the \nufftcf algorithms and implementation choices, while
Appendix~\ref{app:realspace-algos} covers the real-space algorithms.
%
\section{Related Works and Methodology}
\label{sec:related_methodology}
This section is not intended as an exhaustive review of the correlation-function and
spectral-estimation literature for irregularly-sampled data; several dedicated reviews
already serve that purpose (e.g. \cite{Rehfeld2011} for kernel-weighted and
interpolation-based estimators, \cite{BABU2010359} for Fourier-domain spectral-analysis
methods). Rather, our aim here is methodological: we introduce, in the minimal form
needed to motivate and formulate \nufftcf, the four broad families of approaches used in
the literature to handle irregular sampling
(Sections~\ref{sec:slotting}--\ref{sec:lombscargle}), before turning to the Non-Uniform
Fast Fourier Transform (Section~\ref{sec:nufft}) and showing how it can be combined with
kernel-weighted estimators to yield the \nufftcf approach
(Section~\ref{sec:nufftcf_approach}).
\subsection{Correlation estimation on a regular grid: the Wiener--Khinchin theorem}
\label{sec:wk_regular}
Let $x_t$, $t = 0,\dots,n-1$, be a zero-mean, weakly stationary process sampled at $n$
regularly spaced times with sampling interval $\Delta t$. The (biased) sample
autocovariance at lag $k\Delta t$ is
\begin{equation}
\hat R_{xx}(k)=\frac{1}{n}\sum_{t=0}^{n-k-1}x_t\,x_{t+k},
\qquad
k=0,\dots,n-1,
\label{eq:acf_regular}
\end{equation}
and the sample ACF is\footnote{This normalization follows MATLAB convention
(\url{https://uk.mathworks.com/help/matlab/ref/xcorr.html}), where $\hat{R}_{xx}(0)$ is
the total energy of the signal. For cross-correlation, the normalization factor is
$\sqrt{\hat{R}_{xx}(0)\hat{R}_{yy}(0)}$.}
$\hat\rho_{xx}(k)=\hat R_{xx}(k)/\hat R_{xx}(0)$. The sample cross-covariance and
cross-correlation between two co-sampled series $x_t$ and $y_t$ are defined analogously.

For weakly stationary processes, the Wiener–Khinchin theorem establishes that the
autocovariance function and the power spectral density form a Fourier-transform pair.
Similarly, the cross-covariance function and the cross-spectral density also form such a
pair. For finite sampled data, these relationships are numerically realized through the
discrete Fourier transform (DFT): the cross-spectrum is estimated as
\begin{equation}
\hat S_{xy}(f)=\hat X(f)\hat Y^*(f),
\,
\hat R_{xy}(k)=\mathcal F^{-1}\left[\hat S_{xy}(f)\right](k),
\label{eq:wienerkhinchin}
\end{equation}
where $\hat X(f)=\mathcal F[x](f)$ (and similarly for $\hat Y(f)$) denotes the DFT of
the sampled series. On a regular grid, this identity is evaluated efficiently using a
pair of FFTs---a forward FFT to compute $\hat X(f)$ and $\hat Y(f)$, a pointwise
complex-conjugate product to form $\hat S_{xy}(f)$, and an inverse FFT to recover
$\hat R_{xy}(k)$---yielding an $O(n\log n)$ algorithm instead of the $O(n^2)$ direct
lag-domain summation of Equation~\ref{eq:acf_regular}.

This FFT-based approach provides the regular-grid baseline for correlation estimation.
However, for irregularly-sampled data, where the sample times $t_1 < t_2 < \dots < t_n$
are not equally spaced, neither the sum in Equation~\ref{eq:acf_regular} nor the FFT
pair can be applied directly. The literature describes four broad strategies to address
this issue, which we outline below.
\subsection{Slotting and the discrete correlation function}
\label{sec:slotting}
The earliest widely used approach specific to irregular sampling is the
\textit{slotting} technique, introduced in astronomy by \cite{edelson1988} as the
discrete correlation function (DCF). Slotting bins the products of pairs of
(standardized) observations by their time-lag separation into fixed-width bins and
averages within each bin. While simple and model-free, the resulting
correlation-function estimate is not guaranteed to be positive semi-definite and may
require post-processing \citep{Rehfeld2011}. This limitation has motivated the
development of kernel-weighted techniques (Section~\ref{sec:kernel_estimators}).

A refinement of the DCF, proposed by \cite{alexander97} (ZDCF-transform), addresses the
sparse, unevenly sampled light curves typical of AGN monitoring by using
\textit{equal-population binning} rather than fixed-width bins. In this approach, bin
boundaries are chosen adaptively so that each bin contains approximately an equal number
of pairs. Additionally, a Fisher $z$-transform \citep{Fisher1915} is applied to
stabilize the variance of the correlation estimate in each bin before computing
confidence intervals. \pyzdcf \citep{jankov2022pyzdcf} is the reference Python
implementation of ZDCF and remains the standard tool for CCF estimation in AGN
reverberation-mapping studies. Both the DCF and ZDCF require a pairwise scan of the
data, resulting in an $O(n^2)$ computational cost.
\subsection{Kernel-weighted estimators}
\label{sec:kernel_estimators}
Kernel-weighted estimators replace the hard bin assignment of slotting with a smooth
weighting function (kernel) of the time-lag separation. This allows pairs to contribute
to a lag estimate in proportion to the proximity of their separation to that lag. In
this context, \cite{Stoica2009} proposed a $\sinc$ kernel, while \cite{bjornstad2001}
introduced a Gaussian kernel. Notably, the slotting technique of \cite{edelson1988} can
be interpreted as a rectangle (or box) kernel.

The kernel-weighted cross-correlation estimator is defined as:
\begin{equation}
\hat\rho_{xy}(k \Delta\tau) =
\frac{\displaystyle\sum_{i=1}^{n_x}\sum_{j=1}^{n_y} x_i\, y_j \, b_k(d)}
     {\displaystyle\sum_{i=1}^{n_x}\sum_{j=1}^{n_y} b_k(d)},
\label{eq:kernel_estimator}
\end{equation}
where $d = t^y_j - t^x_i - k \Delta\tau$ represents the deviation from the target lag.
Typical choices for $b_k(d)$ include the rectangle/box kernel \citep{edelson1988}:
\begin{equation}
b_k(d) = \begin{cases}
1 & \text{for } |d| \leq h/2, \\
0 & \text{otherwise.}
\end{cases}
\label{eq:rectangle_kernel}
\end{equation}
and the Gaussian kernel \citep{bjornstad2001}, which is the primary choice in \pastas
and \nufftcf:
\begin{equation}
b_k(d) = \frac{1}{\sqrt{2\pi}\,h}\exp\left(-\frac{d^2}{2h^2}\right).
\label{eq:Gaussian_kernel}
\end{equation}
Here, $h$ is a bandwidth parameter typically set to match the mean sampling interval
$\Delta t^{xy}$, see \cite{Rehfeld2011} for details. Another typical choice, not yet
implemented in the current version of \nufftcf, is the Hanning kernel
\citep{Max-Moerbeck2014}.

Given the irregular sampling common in geoscientific time series, \cite{Rehfeld2011}
conducted a systematic benchmark of kernel-based estimators (Gaussian, $\sinc$,
rectangle) against linear interpolation and the Lomb--Scargle approach, the latter being
addressed in the following section. Using synthetic sinusoidal and autoregressive
processes with controlled sampling irregularity, they found that the Gaussian-kernel
estimator consistently exhibited the lowest, or close to the lowest, root-mean-square
error (RMSE) and bias across nearly all test cases. This led the authors to recommend it
over linear interpolation for irregularly sampled data.

This methodology underpins the ACF estimator implemented in \pastas
\citep{collenteur2019}, a widely used Python package for groundwater time-series
analysis. \pastas offers Gaussian, rectangle, and regular-grid binning options for its
ACF. As with slotting, kernel-weighted estimators evaluate a direct pairwise sum, with
an overall cost of $O(n^2)$ for a full lag range, as measured for \pastas in
Section~\ref{sec:benchmarks}. This is not intrinsic to the kernel-weighted definition:
when the sample times are sorted, a two-pointer scan brings the cost down to $O(n)$ per
lag, hence $O(nK)$ for $K$ requested lags. The real-space algorithms of
Appendix~\ref{app:realspace-algos} implement this scan, and the timing benchmarks of
Section~\ref{sec:benchmarks} confirm the resulting linear scaling in $n$ at fixed $K$.
\subsection{Lomb--Scargle-based (Fourier-domain) estimation}
\label{sec:lombscargle}
A conceptually different approach, introduced by
\cite{1976Ap&SS..39..447L,1982ApJ...263..835S} (see also
\cite{1989ApJ...343..874S,2018ApJS..236...16V}), estimates the cross-correlation
function by first computing the cross-spectrum via the Lomb--Scargle Fourier transform
(LSFT). The LSFT is an irregular-sampling generalization of the DFT built from a
least-squares sinusoid fit. The (cross-)periodogram is then formed, and an inverse
Fourier transform is applied to the result, effectively applying the Wiener--Khinchin
theorem in the Fourier domain rather than directly binning or weighting observation
pairs in the lag domain. More broadly, the LSFT is one instance of a wider family of
Fourier-domain approaches to spectral analysis of nonuniformly-sampled data,
comprehensively reviewed by \cite{BABU2010359}.

\cite{Rehfeld2011} found this approach competitive with kernel methods for univariate
(ACF) problems but markedly worse for bivariate (CCF) problems. While the Lomb--Scargle
method reuses the transform-multiply-invert idea of Section~\ref{sec:wk_regular}, it
replaces the FFT with a nonuniform transform tailored to irregular sampling, with a
computational cost of $O(n\,n_f)$ for $n_f$ frequencies. In practice, this transform can
itself be evaluated efficiently via a non-uniform FFT (NUFFT), following the approach of
\cite{Garrison2024}, which uses the same NUFFT library (FINUFFT) as \nufftcf, as
described in the next section.
\subsection{The Non-Uniform Fast Fourier Transform}
\label{sec:nufft}
The NUFFT computes Fourier sums between nonuniformly and uniformly spaced points without
forming the $O(n\cdot m)$ direct sum over all $n$ sample points and $m$ output
frequencies (or vice versa). Two of its three standard types are relevant here (see sign
convention in Section~\ref{app:finufft-notes}):
\begin{itemize}
    \item \textbf{type~1} (``nonuniform to uniform''): given values $x_j$ at nonuniform
    points $t_j$, $j=1,\dots,n$, compute
    \begin{equation}
    \hat X(f_k) = \sum_j x_j\, e^{i f_k t_j} := X_k
    \label{eq:nufft_type1}
    \end{equation}
     at $m$ \textit{uniform} output frequencies $f_k$.
    \item \textbf{type~2} (``uniform to nonuniform''): the adjoint operation, evaluating
    a uniformly-gridded Fourier series at arbitrary nonuniform output points:
        \begin{equation}
    	x_j = \sum_k X_k\, e^{-i f_k t_j}
    	\label{eq:nufft_type2}
    \end{equation}
    
\end{itemize}

Modern NUFFT algorithms, such as \texttt{FINUFFT} \citep{barnett2019finufft} (used by
\nufftcf), compute both types in $O((n+m)\log(1/\varepsilon))$ time, where $\varepsilon$
is the requested accuracy. This is achieved by (i) \textit{gridding}, where each
nonuniform sample is spread onto a fine uniform grid using a compactly-supported
interpolation kernel, (ii) taking a standard FFT of the fine grid, and (iii)
\textit{deconvolving} by dividing by the Fourier transform of the spreading kernel to
correct for the smoothing introduced by the kernel. Type~2 reverses this sequence.

\texttt{FINUFFT} uses the ``exponential of semicircle''
kernel\footnote{$\phi_\beta(x)=e^{\beta\sqrt{1-x^2}}$ for $|x|\leq 1$ ($0$ elsewhere).}
as preading kernel, which is chosen to first minimize aliasing error for a given support
width and secondly because its Fourier transform can be computed easily using a
Gauss-Legendre quadrature integration. Such spreading kernel is an internal
numerical-accuracy device of the NUFFT algorithm itself, distinct from the statistical
weighting kernels $b_k(\cdot)$ of Equation~\ref{eq:Gaussian_kernel}. We return below to
the distinction between the two kernels (spreading and statistical weighting).
\subsection{From kernel-weighted estimators to NUFFT: the nufftcf approach}
\label{sec:nufftcf_approach}
\nufftcf central observation is that the kernel-weighted estimator of
Equation~\ref{eq:kernel_estimator} can be evaluated via Wiener--Khinchin using a NUFFT,
rather than via a direct pairwise sum, whenever the weighting kernel $b_k(\cdot)$ is
itself expressible as (or well approximated by) a smooth, compactly supported function
of the lag---which the Gaussian and rectangle kernels used by \pastas are. Concretely,
\nufftcf:
\begin{enumerate}
    \item forms the (irregularly sampled) power or cross-spectrum of the input series
    with a \textbf{type~1 NUFFT} onto a uniform frequency grid whose resolution is set
    by the requested lag range and kernel bandwidth;
    \item evaluates the correlation estimate at the requested lags by inverting this
    spectrum with a \textbf{type~2 NUFFT}, applying the Wiener--Khinchin theorem in the
    same spirit as the regular-grid FFT pair of Section~\ref{sec:wk_regular}, but
    replaced by their nonuniform-capable counterparts.
\end{enumerate}
Algorithms~\ref{alg:ccf-nufft}--\ref{alg:acf-nufft} in Appendix~\ref{app:algorithms}
give the complete pseudocode of the CCF and ACF estimators, respectively, and
Appendix~\ref{app:finufft-notes} details the implementation choices dictated by
\texttt{FINUFFT} conventions (domain mapping, sign convention, oversampling factor
$N_1$, and the interior-lag normalization used to avoid an edge-smoothing bias). The
cost of evaluating the correlation function at $K$ lags from $n$ (irregular) samples has
two parts. The spectral part (two type~1 and one type~2 NUFFT calls) costs
$O(n\log n+K)$ for $N_1\propto n$. The normalization by the effective pair count,
described below, costs $O(n)$ \emph{per lag}, hence $O(nK)$. For a fixed $K$ the two
terms are hardly distinguishable over one or two decades in $n$, and $n\log n$ overtakes
$nK$ only for $\ln n\gtrsim K$, out of reach for any realistic $K$: in practice the
NUFFT estimators scale as $O(nK)$, i.e.\ linearly in $n$ at fixed $K$
(Section~\ref{sec:benchmarks} gives the measured coefficients). This remains far below
the measured $O(n^2)$ of the pairwise bin methods. Because this route implicitly
represents the signal on a finite, periodic Fourier basis, it is subject to a small
spectral-leakage bias for strongly periodic input with irregular or gappy sampling
(quantified in Section~\ref{sec:benchmarks} and controllable via the number of Fourier
modes). \nufftcf therefore also provides a direct, ``real-space'' evaluation of the same
kernel-weighted estimator. It relies on the same two-pointer scan, applied to both the
numerator and the denominator, hence also costs $O(nK)$, but without any periodicity
assumption: it is the accuracy reference, the recommended choice when this bias must be
avoided entirely, and the faster one when only a few lags are requested, the NUFFT
estimator winning beyond a kernel-dependent number of lags $K^*$
(Sections~\ref{sec:benchmarks} and~\ref{sec:discussion}).

\textbf{Effective pair count.} Every estimate in Equation~\ref{eq:kernel_estimator} is
normalized by $b_k(\cdot)$ summed over all pairs---the \textit{effective number of
pairs} contributing to lag $k$, which both normalizes the correlation estimate and flags
under-sampled lags. Computed naively this denominator is itself an $O(n^2)$ all-pairs
sum. Because the sample times are sorted, \nufftcf instead computes it in $O(n)$ per lag
with a two-pointer scan implemented as a \texttt{numba}-compiled \citep{Numba2015}
kernel per supported weighting kernel
(Algorithms~\ref{alg:b-generic}--\ref{alg:b-rect-generic} in
Appendix~\ref{app:pair-count}), and, for the regular-grid classic-FFT estimators,
recovers the same quantity for free by filtering the deterministic raw-pair-count ramp
with the same discrete filter applied to the correlation numerator. This avoids
reintroducing an $O(n^2)$ step per lag; the remaining $O(n)$ per lag nevertheless makes
this scan a major share of the cost of the NUFFT estimators (dominant for the Gaussian
kernel, Section~\ref{sec:benchmarks}), whose overall scaling is thus $O(nK)$ rather than
$O(n\log n)$.

\textbf{Unified API.} All three estimator families implemented in
\nufftcf---NUFFT-based, real-space, and classic-FFT (regular grid only)---share the
calling convention \texttt{fn(lags, tx, x, [ty, y], ...) -> (c,
b)}\footnote{\texttt{(c,b)} standing for "correlation estimate" and "effective pair
count".}, so that a user can move between them, e.g.\ to check a NUFFT-based result
against the real-space reference on a subset of the data, without altering the rest of
an analysis pipeline.
\section{Comparison with pastas and pyZDCF}
\label{sec:comparison}
We now validate \nufftcf against two widely used, independently implemented estimators
--- \pastas (kernel-weighted, real-space, Section~\ref{sec:kernel_estimators}) and
\pyzdcf (equal-population slotting, Section~\ref{sec:slotting}) --- on synthetic series
for which the ``true'' correlation function is known analytically. We treat ACF and CCF
separately, since \pastas does not expose a cross-correlation estimator equivalent to
its ACF\footnote{An internal cross-correlation code exists from which the ACF is
computed, but it is not part of its public API. In the \texttt{pastas/pastas-plugins}
repository, the cross-correlation code uses the \texttt{scipy.signal.correlate} function
\citep{scipy2020} with the FFT method, which is only valid for equidistant time steps.},
so the CCF benchmark below compares only \nufftcf against \pyzdcf.
\subsection{Test series and experimental setup}
\label{sec:comparison_setup}
\begin{figure}[h]
	\centering
	\includegraphics[width=\columnwidth]{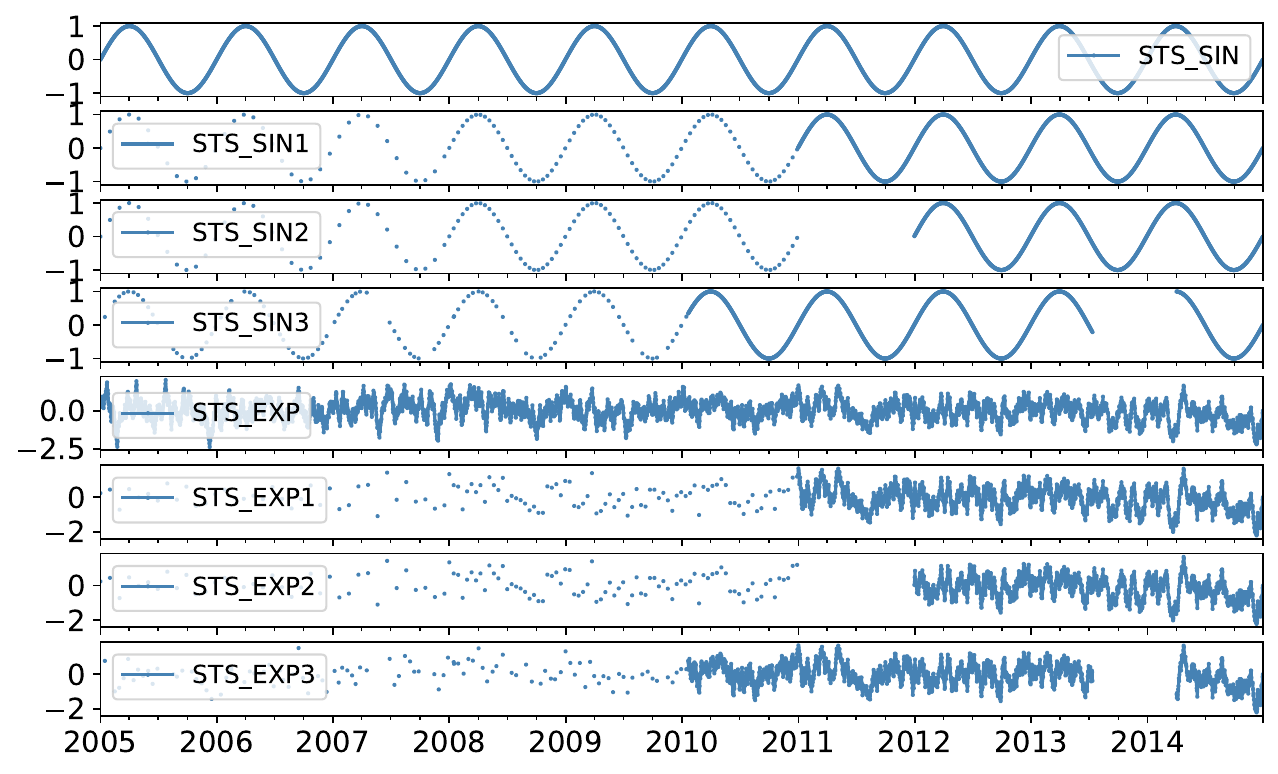}
	\caption{The synthetic series used in the ACF benchmark: the
		fully-sampled reference processes (STS\_SIN, STS\_EXP) and their
		three irregularly-sampled derivatives each (STS\_SIN1--3,
		STS\_EXP1--3), reproduced from the \pastas autocorrelation-benchmark
		tutorial.}
	\label{fig:synthetic_series}
\end{figure}
\textbf{ACF benchmark series\footnote{see \texttt{pastas\_vs\_nufftcf.ipynb} and
\texttt{zdcf\_vs\_nufftcf.ipynb} notebooks.}.} We reuse the synthetic
irregularly-sampled series from the \pastas autocorrelation-benchmark
tutorial\footnote{\url{https://pastas.readthedocs.io/stable/benchmarks/autocorrelation.html}},
shown in Figure~\ref{fig:synthetic_series}. Two underlying stationary processes are
generated over a ten-year span at a one-day base resolution:
\begin{itemize}
	\item \texttt{STS\_SIN}: a pure sine wave of period $T=365.25$~d, with analytical
	autocorrelation $\rho(\tau)=\cos(2\pi\tau/T)$;
	\item \texttt{STS\_EXP}: an AR(1)-like process obtained by convolving white noise with
	a causal exponential kernel $e^{-t/\alpha}$, $\alpha=10$~d; because this kernel
	coincides with the process autocovariance, its analytical ACF is
	$\rho(\tau)=e^{-\tau/\alpha}$.
\end{itemize}
Each process is then resampled onto three irregular time grids of different sparsity,
yielding six benchmark series (\texttt{STS\_SIN}1--3, \texttt{STS\_EXP}1--3): (1) a mix
of 30-day and 14-day regular sub-sampling with dense daily coverage over the last four
years; (2) the same scheme with an additional three-year data gap; and (3) re-indexing
onto a set of real, highly irregular groundwater-monitoring dates spanning 1960--2015,
used asis from the \pastas tutorial. Series~3 is by far the sparsest and most irregular
of the three ($n\sim 500\-600$ points over 55~years with long gaps), while series~1--2
alternate between decades of sparse sampling and a few years of dense daily coverage.

\textbf{CCF benchmark series\footnote{see \texttt{nufftcf\_ccf\_demo.ipynb} notebook.}.}
Two complementary CCF setups are used, both without access to \pastas:
\begin{itemize}
	\item Shifted-timestamp \texttt{STS\_SIN/EXP} series: The same \texttt{STS\_SIN} and
	\texttt{STS\_EXP} processes used in the ACF benchmark are drawn over
	$n_{\rm days}=3650$~d, and two independent, irregular observation series $x(t_x)$ and
	$y(t_y)$ sample the identical underlying process values, with $y$-timestamps shifted by
	a known delay $\tau_0$ relative to $x$-timestamps ($\sim43\%$ of days retained in each
	series). Because $x$ and $y$ are the same signal read at shifted times, the CCF should
	peak exactly at $\tau=\tau_0$ with the ACF own peak value ($1$ for these unit-variance
	processes). We use $\tau_0=60$~d as the fiducial case (Figure~\ref{fig:ccf_sin_exp})
	and additionally sweep $\tau_0\in\{20,60,120\}$~d for the \texttt{STS\_EXP} series
	(Figure~\ref{fig:ccf_tau0}) to check that the recovered peak location tracks the
	injected delay.
	\item Coupled Ornstein--Uhlenbeck (OU) processes with variable $\rho$. Two independent
	latent OU-like processes\footnote{Concretely, each process is built by causally
	filtering unit-variance white noise $\eta_t\sim\mathcal N(0,1)$ through a discrete
	exponential kernel, $e_t \propto \sum_{k\geq 0} e^{-k/\alpha}\,\eta_{t-k}$, truncated
	after a $200$-day burn-in (negligible relative to $\alpha=10$~d) and rescaled to unit
	variance. Since $e^{-k/\alpha}$ is geometric, this sum telescopes exactly into the
	AR(1) recursion $e_t = \phi\, e_{t-1} + \eta_t$ with $\phi=e^{-1/\alpha}$. This is the
	exact (not Euler--Maruyama) discretisation of the OU Stochastic Differential Equation
	$dX = -\theta X\,dt + \sigma\,dW$, $\theta=1/\alpha$, at unit time step, with
	innovation variance $\sigma^2(1-e^{-2\theta})/(2\theta)=1$,
	i.e.\ $\sigma=\sqrt{2\theta/(1-e^{-2\theta})}$, so that $\eta_t$ matches the
	unit-variance driving noise above. The resulting process is stationary with
	autocorrelation $e^{-|\tau|/\alpha}$ up to the burn-in transient, which is what
	underlies Equation~\eqref{eq:ou_theory_ccf}.} $e_1(t)$, $e_2(t)$ are generated with the
	same exponential-kernel construction as \texttt{STS\_EXP} ($\alpha=10$~d), and combined
	as
	\begin{align}
		X_1(t) &= e_1(t), \nonumber \\
		X_2(t) &= \rho\, e_1(t-\tau_0) + \sqrt{1-\rho^2}\, e_2(t),
		\label{eq:ou_coupling}
	\end{align}
	with $\tau_0\!=\!\text{60\,d}$ and $\rho\!\in\!\{0.1,0.2,0.3,0.5,0.7,0.95\}$. Unlike
	the shifted-timestamp case, here the lag is \emph{physical}: $X_2$ is a genuinely
	delayed, partially decorrelated copy of $X_1$, contaminated by an independent noise
	process. The two series are sampled on independent irregular grids retaining $\sim40\%$
	and $\sim60\%$ of days, respectively, with no artificial timestamp shift. The
	analytical cross-correlation function for this construction is
	\begin{equation}
		\rho_{X_1X_2}(\tau) = \rho\, e^{-|\tau-\tau_0|/\alpha},
		\label{eq:ou_theory_ccf}
	\end{equation}
	i.e.\ an exponential peak of height $\rho$ centered at $\tau=\tau_0$
	(Figure~\ref{fig:ccf_ou}).
\end{itemize}

\textbf{Software configuration.} For \nufftcf we evaluate both the rectangle and
Gaussian kernels (Equations~\ref{eq:rectangle_kernel}, \ref{eq:Gaussian_kernel}) via the
NUFFT-based estimator (and, for the fiducial CCF case, the direct real-space evaluation
as an accuracy cross-check), using \texttt{bin\_width}$=0.5$~d as the kernel-bandwidth
control parameter for all ACF and CCF runs, and evaluating every integer lag from $1$ to
$365$~d (ACF) or $1$ to $180$~d (CCF).

For \pastas we call \texttt{pastas.stats.acf} with \texttt{bin\_method} set to
\texttt{"rectangle"} and \texttt{"Gaussian"} and \texttt{max\_gap}=$30$~d, over the same
lag range.

For \pyzdcf we use \texttt{uniform\_sampling = False}, \texttt{omit\_zero\_lags = True},
and its equal-population bin-size parameter \texttt{minpts}, the closest analogue to
\nufftcf and \pastas bandwidth: we report $\mathrm{minpts}=11$ (its default in
practice\footnote{In practice, the default $\mathrm{minpts}=0$ in \pyzdcf is internally
replaced by $\mathrm{ENOUGH}=11$ (see \texttt{pyzdcf/pyzdcf.py}).}) for all CCF runs and
both $\mathrm{minpts}=11$ and $\mathrm{minpts}=25$ for the ACF runs, with asymmetric
Monte-Carlo error bars from $50$ (ACF) or $100$ (CCF) resamplings. Because \pyzdcf
adapts its bin boundaries to the pair count rather than to a fixed lag spacing, the
number of lags it returns is set by the data and by $\mathrm{minpts}$, i.e., not chosen
by the user. It therefore always produces \emph{far fewer} lag points than the fixed
grid used by \nufftcf, typically resulting in $\sim100-500$ bins for the ACF series
here, compared to the full requested grid of $365$ or $180$ points for \nufftcf. This is
a direct consequence of \pyzdcf equal-population-per-bin design
(Section~\ref{sec:slotting}).
\subsection{Autocorrelation: \nufftcf vs \pastas and \pyzdcf}
\label{sec:comparison_acf}
\begin{figure*}[p]
	\centering
	\includegraphics[width=0.85\textwidth]{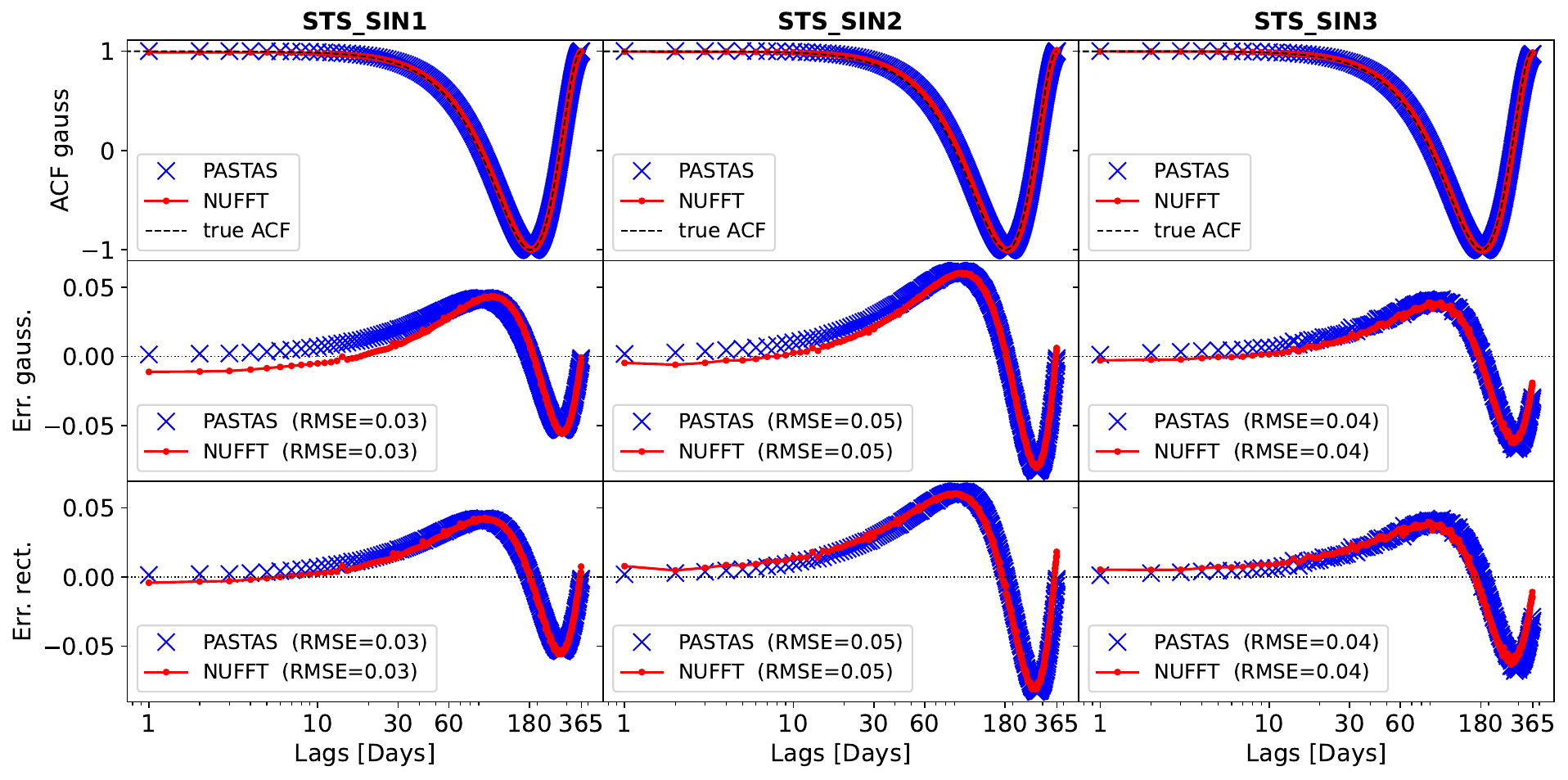}
	\caption{ACF of the\texttt{STS\_SIN}1--3 series: signal (top row), \pastas
		vs.\ \nufftcf Gaussian-kernel ACF against the analytical truth
		(second row), and residuals for the Gaussian and rectangle kernels
		(third and fourth rows), with matching RMSE values for both
		estimators in every panel.}
	\label{fig:pastas_sin}
\end{figure*}

\begin{figure*}
	\centering
	\includegraphics[width=0.85\textwidth]{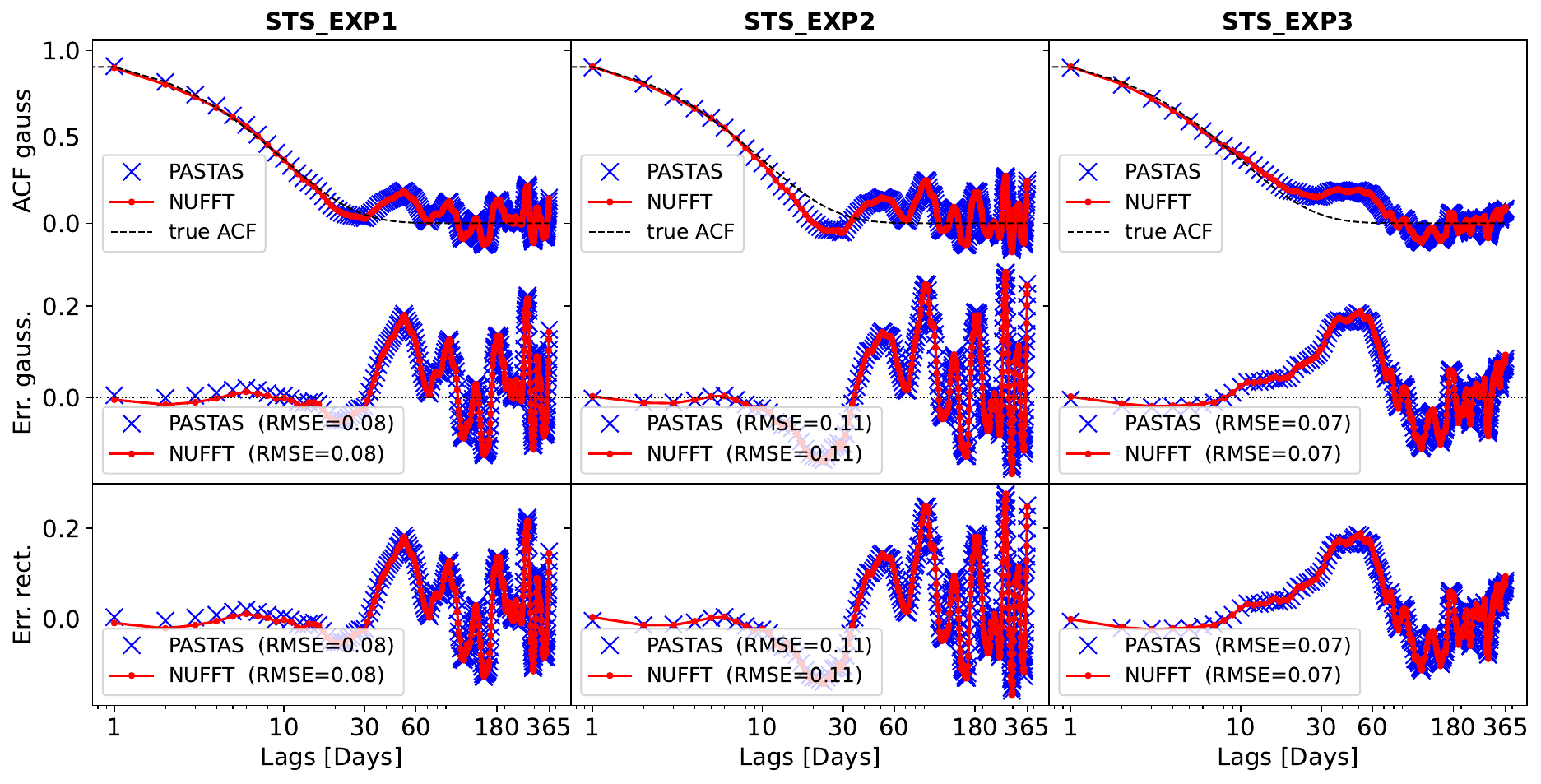}
	\caption{Same as Figure~\ref{fig:pastas_exp}, for the \texttt{STS\_EXP}1--3
		series.}
	\label{fig:pastas_exp}
\end{figure*}

Figures~\ref{fig:pastas_sin}--\ref{fig:pastas_exp} compare \nufftcf and \pastas on the
\texttt{STS\_SIN} and \texttt{STS\_EXP} series, respectively, showing the estimated ACF
together with the residual to the analytical truth for both the Gaussian and rectangle
kernels. Across all six series the two implementations are visually indistinguishable
and their Root Mean Square Error (RMSE) against the analytical ACF agree to within the
reported precision (e.g. $\mathrm{RMSE}=0.03/0.05/0.04$ for \texttt{STS\_SIN}1/2/3 and
$\mathrm{RMSE}=0.08/0.11/0.07$ for \texttt{STS\_EXP}1/2/3, identical for \pastas and
\nufftcf in every case). This confirms that \nufftcf, NUFFT-based Wiener--Khinchin
methodology, reproduces the direct pairwise kernel-weighted estimator of
Equation~\ref{eq:kernel_estimator} to the accuracy expected from its spectral-leakage
bias (Section~\ref{sec:nufftcf_approach}), which here is negligible compared to the
sampling and estimation noise common to both methods. The residuals for both methods
grow with $n$-dependent gaps and departures from stationarity (e.g.\ around the
transition between sparse and dense sampling in \texttt{STS\_EXP}1/2 near lag
$\sim$40--100~d), which is a property of the underlying data and kernel bandwidth rather
than of either estimator.

\begin{figure*}
	\centering
	\includegraphics[width=0.85\textwidth]{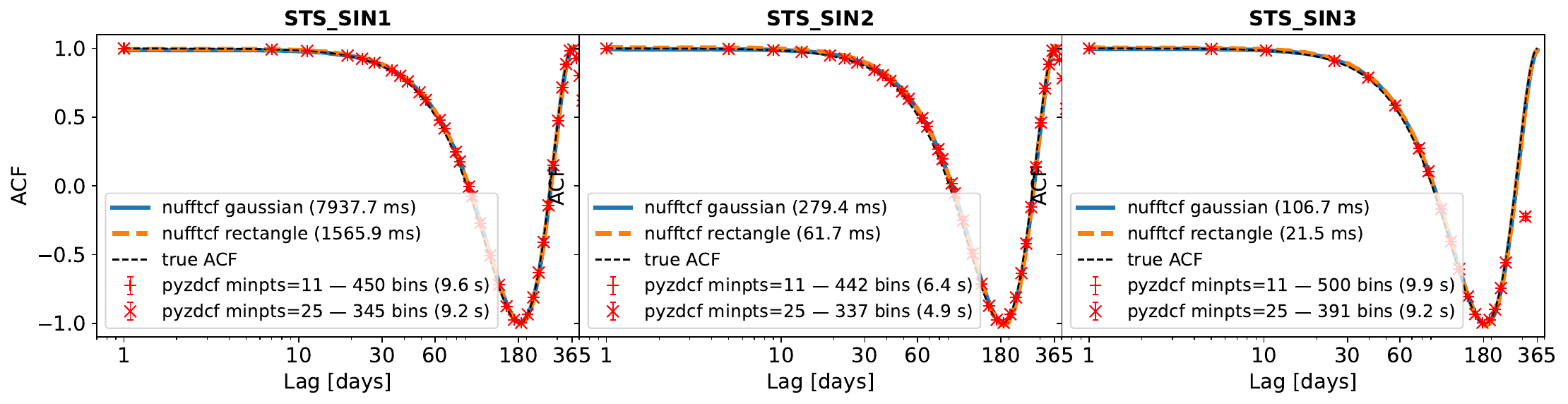}
	\includegraphics[width=0.85\textwidth]{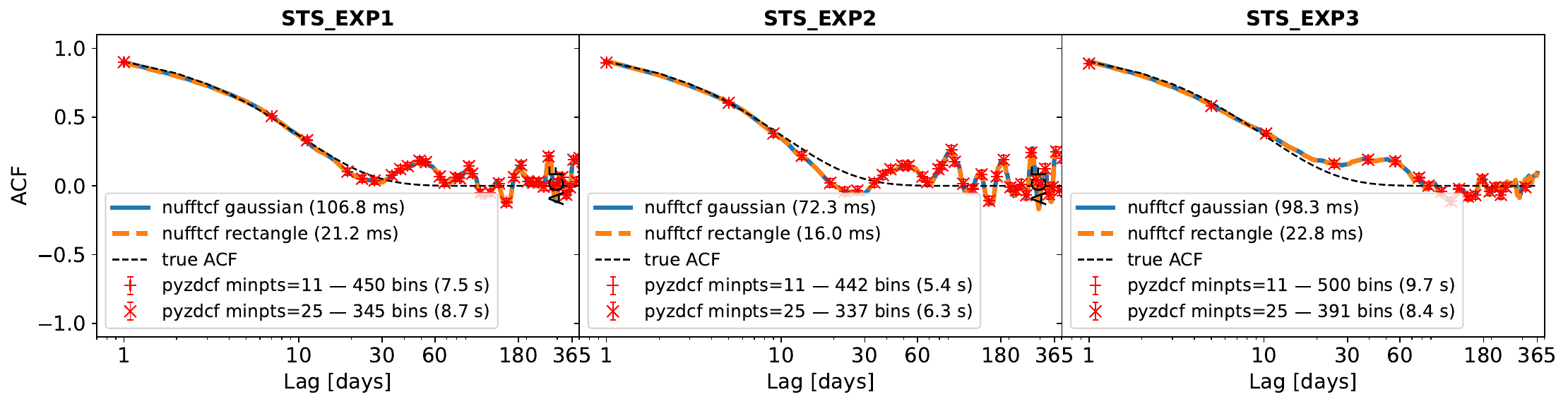}
	\caption{Top: ACF of the \texttt{STS\_SIN}1--3 series: signal (top row) and ACF
		(bottom row), comparing \nufftcf (Gaussian and rectangle kernels,
		with per-series computation time) to \pyzdcf (\texttt{minpts}=11
		and 25, with number of returned bins and computation time) and to
		the analytical truth. Bottom: Same for the \texttt{STS\_EXP}11--3
		series.}
	\label{fig:zdcf_sin_exp}
\end{figure*}

Figure~\ref{fig:zdcf_sin_exp} repeats the comparison against \texttt{pyzdcf}. The
\texttt{nufftcf} Gaussian and rectangle estimates, which closely follow the analytical
ACF, agree with the \texttt{pyzdcf} points (within their Monte-Carlo error bars)
wherever the two methods are directly comparable. The main qualitative differences are
in resolution and computational cost: \texttt{nufftcf} returns a value at every
requested lag for these series (e.g., $365$ points) in at most a few hundred
milliseconds, whereas \texttt{pyzdcf} returns only $\sim 300$--$500$ equal-population
bins (even fewer for the coarser $\mathrm{minpts}=25$ setting) and requires several
seconds per series. These timing differences are driven by \texttt{pyzdcf} $O(n^2)$
pairwise scan and Monte Carlo error resampling, compared to \texttt{nufftcf} $O(nK)$
cost (NUFFT evaluation plus pair-count normalization, with a small prefactor;
Section~\ref{sec:benchmarks}).
\subsection{Cross-correlation: nufftcf vs pyZDCF (\texttt{STS\_SIN/EXP} series)}
\label{sec:comparison_ccf}
\begin{figure*}
	\centering
	\includegraphics[width=0.85\textwidth]{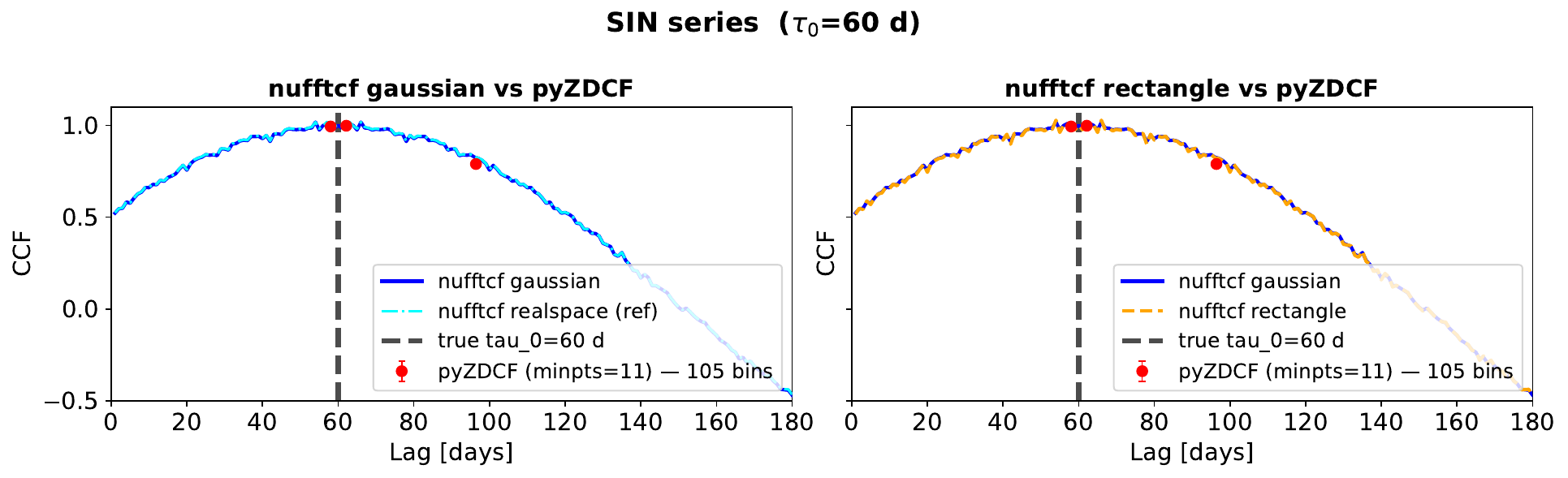}
	\includegraphics[width=0.85\textwidth]{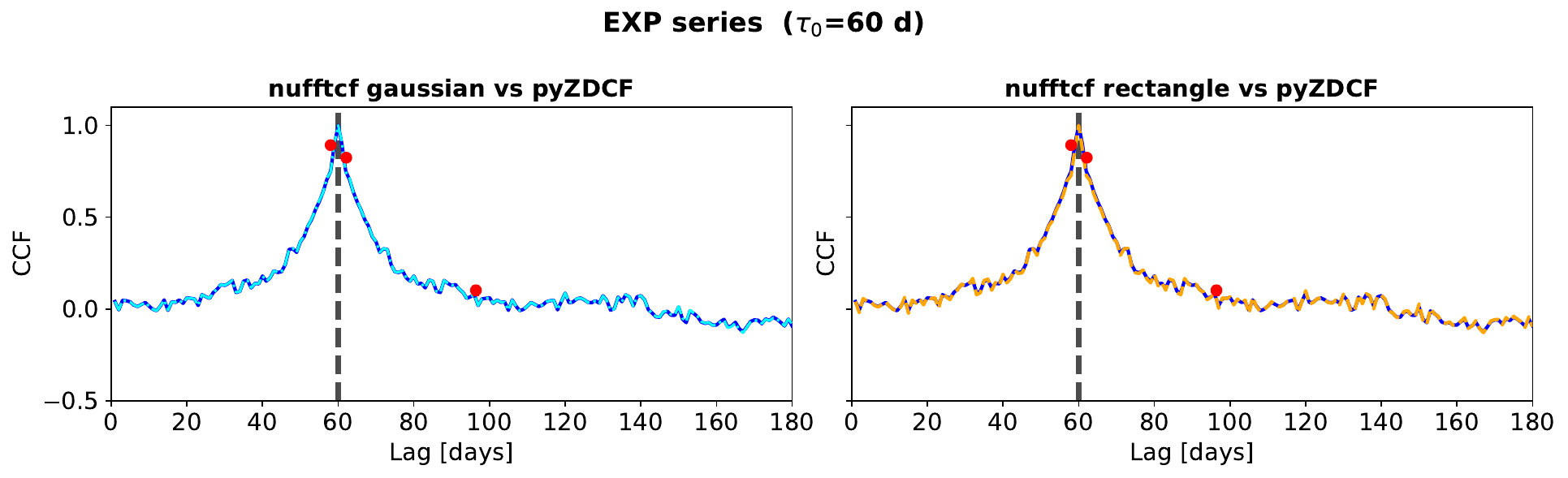}
	\caption{Top: CCF of the shifted-timestamp  \texttt{STS\_SIN} test
		($\tau_0=60$~d): \nufftcf Gaussian vs. its own real-space reference
		(left) and vs. its rectangle kernel (right), both compared to
		\pyzdcf. Bottom: Same legend for the shifted-timestamp
		\texttt{STS\_EXP} test.}
	\label{fig:ccf_sin_exp}
\end{figure*}

\begin{figure*}
	\centering
	\includegraphics[width=0.85\textwidth]{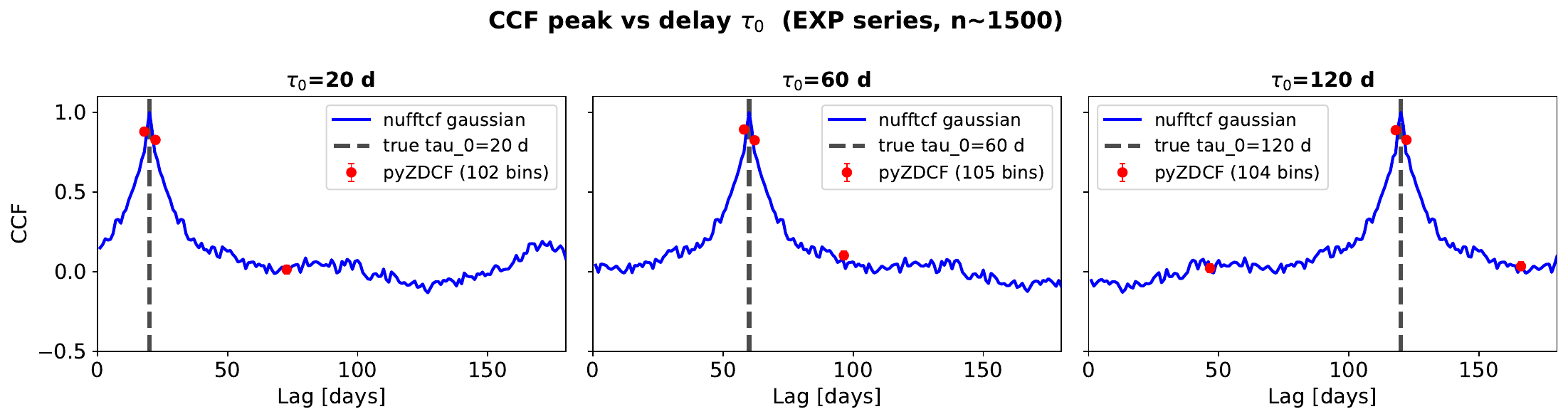}
	\caption{CCF peak recovery for the shifted-timestamp EXP network
		test as the injected delay $\tau_0$ varies over $\{20,60,120\}$~d
		($n\sim1500$ points per network): \nufftcf Gaussian vs.\ \pyzdcf,
		with the true delay marked by the vertical dashed line.}
	\label{fig:ccf_tau0}
\end{figure*}

Figure~\ref{fig:ccf_sin_exp} shows the fiducial $\tau_0=60$~d shifted-timestamp case for
the \texttt{STS\_SIN} and \texttt{STS\_EXP} series, respectively. The \nufftcf
Gaussian-kernel estimate is compared against its real-space reference (left panels) and
its rectangle-kernel estimate (right panels). In both panels, the \pyzdcf estimate is
also displayed. The three \nufftcf variants agree essentially exactly with one another,
and all three also agree with the \pyzdcf sparser set of binned points within its error
bars. Additionally, they correctly locate the CCF peak at the injected delay
$\tau=\tau_0$. Figure~\ref{fig:ccf_tau0} confirms that this peak recovery is robust as
$\tau_0$ is varied over $\{20, 60, 120\}$~d, exemplified for the \texttt{STS\_EXP}
series: both \nufftcf and \pyzdcf recover peaks consistent with the true delay (marked
by the vertical dashed line), with the \nufftcf continuous curve making the peak
location and shape considerably easier to read off than the \pyzdcf handful of bins per
panel.

\begin{figure*}
	\centering
	\includegraphics[width=0.85\textwidth]{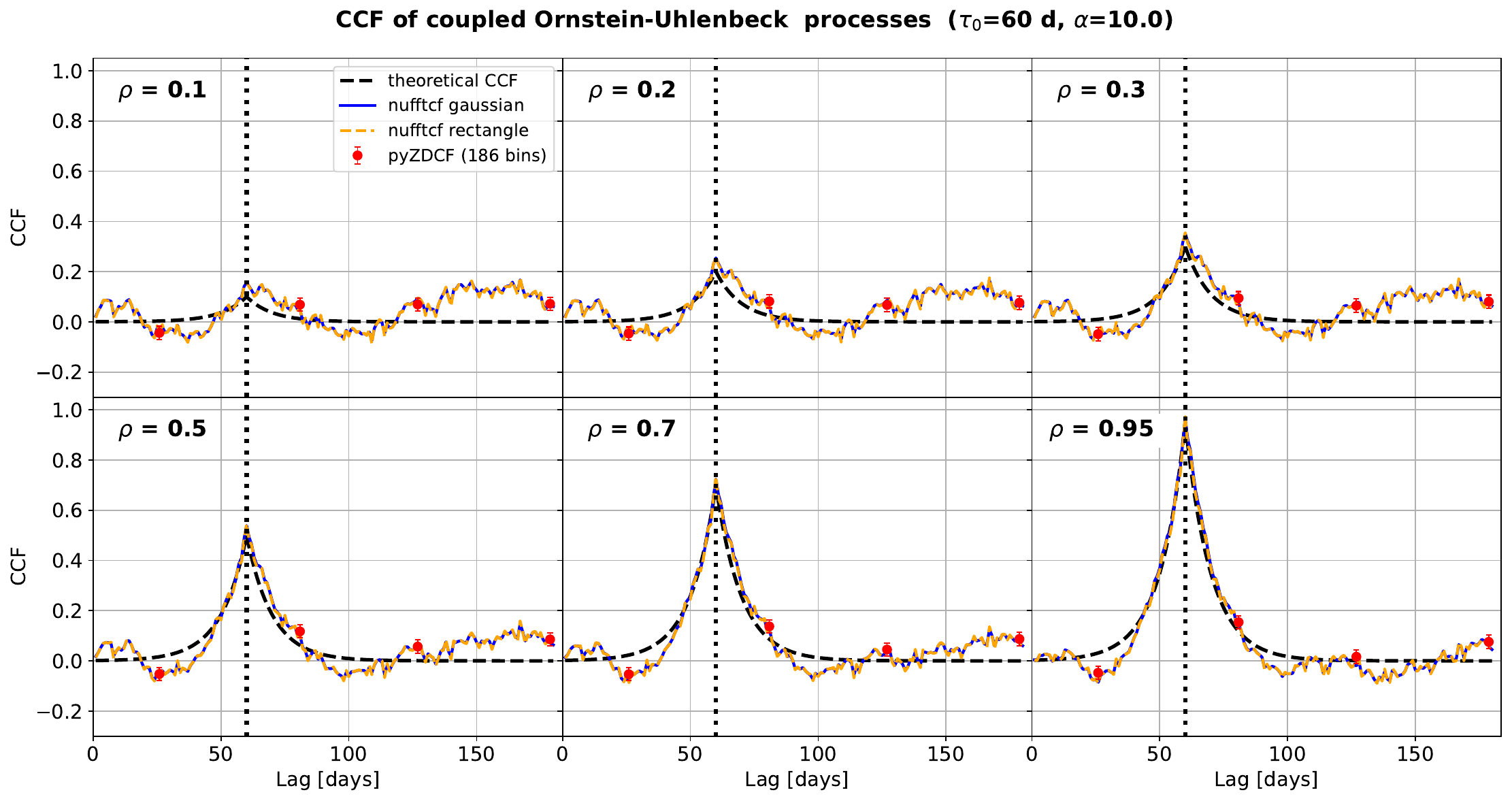}
	\caption{CCF of the coupled Ornstein--Uhlenbeck processes of
		Equation~\ref{eq:ou_coupling} ($\tau_0=60$~d, $\alpha=10$~d), for
		coupling strengths $\rho=0.1,0.4,0.7,0.95$: \nufftcf (Gaussian and
		rectangle kernels) and \pyzdcf against the analytical CCF of
		Equation~\ref{eq:ou_theory_ccf}.}
	\label{fig:ccf_ou}
\end{figure*}
\subsection{Cross-correlation: nufftcf vs pyZDCF (coupled OU-processes)}
\label{sec:comparison_ccf_ouproc}
\begin{table*}
	\centering
	\caption{Comparison of cross-correlation peaks for different values of $\rho$.}
	\label{tab:peak_comparison_transposed}
	\resizebox{0.9\textwidth}{!}{%
		\begin{tabular}{l cc cc cc cc cc cc}
			\toprule
			$\tau_0=60$~d & \multicolumn{2}{c}{$\rho=0.1$} & \multicolumn{2}{c}{$\rho=0.2$} & \multicolumn{2}{c}{$\rho=0.3$} & \multicolumn{2}{c}{$\rho=0.5$} & \multicolumn{2}{c}{$\rho=0.7$} & \multicolumn{2}{c}{$\rho=0.95$} \\
			\cmidrule(lr){2-3}\cmidrule(lr){4-5}\cmidrule(lr){6-7}\cmidrule(lr){8-9}\cmidrule(lr){10-11}\cmidrule(lr){12-13}
			Method & val. & loc. & val. & loc. & val. & loc. & val. & loc. & val. & loc. & val. & loc. \\
			\midrule
			\nufftcf Gaussian  & 0.167 & 167 & 0.255 & 60 & 0.351 & 60 & 0.539 & 60 & 0.724 & 60 & 0.962 & 60 \\
			\nufftcf rectangle & 0.178 & 167 & 0.258 & 60 & 0.355 & 60 & 0.546 & 60 & 0.733 & 60 & 0.973 & 60 \\
			realspace Gaussian & 0.181 & 167 & 0.256 & 60 & 0.352 & 60 & 0.538 & 60 & 0.722 & 60 & 0.957 & 60 \\
			\pyzdcf             & 0.072 & 178.7 & 0.082 & 80.9 & 0.094 & 80.9 & 0.117 & 80.9 & 0.138 & 80.9 & 0.154 & 80.9 \\
			\bottomrule
		\end{tabular}
	}
\end{table*}
	
Figure~\ref{fig:ccf_ou} presents the physically coupled Ornstein-Uhlenbeck process test
of Equation~\ref{eq:ou_coupling}, for which \nufftcf (Gaussian and rectangle kernels)
and \pyzdcf estimates are displayed as weel as the analytical CCF
(Equation~\ref{eq:ou_theory_ccf}) as a dashed black curve.
Table~\ref{tab:peak_comparison_transposed} compares the estimated CCF peak (amplitude
and location) across the six tested values of $\rho$, for the three estimators. The
theoretical expectation is a peak value of $\rho$ and a location of $\tau_0=60$~d,
respectively (Eq.~\ref{eq:ou_theory_ccf}). We discuss the two peak charactireistics
recovery.
\subsubsection{Peak location}
For $\rho \geq 0.2$, both \nufftcf kernel variants and the real-space Gaussian estimator
correctly recover $\tau_0=60$~d, whereas \pyzdcf systematically places the peak near
$80.9$~d, i.e.\ a constant offset of about $+21$~d. This offset is an intrinsic
consequence of the equal-population slotting algorithm: the reported lag is the mean lag
of the pairs contained in the highest-correlation bin, not a fixed grid point, and bin
edges are set purely by pair count rather than by physical lag. A test done by retaining
$50\%$ of both series (Equation~\ref{eq:ou_coupling}) shows that \nufftcf recovers the
peak location, while \pyzdcf increases the deviation by $+78.1$. \pyzdcf
equal-population slotting can degrade in an essentially uncontrolled way once the local
pair density near $\tau_0$ (set jointly by \texttt{minpts}, the sampling retention
fractions, and their overlap structure) becomes insufficient to support a bin localized
to the true peak neighborhood. We do not undertake further a detailed study of \pyzdcf
behavior as we focus ourselves on \nufftcf results.

The $\rho=0.1$ case is atypical for all estimators: the peak is found at
$\mathrm{lag}=167$~d (\nufftcf and real-space Gaussian alike) or $178.7$~d (\pyzdcf),
far from $\tau_0=60$~d. At this weak coupling level, the signal-to-noise ratio of the
theoretical CCF is too low for the physical peak to reliably dominate the estimated
cross-correlation function, so the global maximum is instead captured by a noise
fluctuation rather than by the expected physical peak. As a further check of this
interpretation, we repeated the coupled OU-process experiment with the total baseline
extended by a factor of ten (from $3650$ to $36\,500$~d, i.e. ten times more sampled
points at comparable cadence), keeping all other parameters unchanged. With this larger
sample, \nufftcf (both kernels) and the real-space Gaussian estimator recover
$\mathrm{lag}=\tau_0=60$~d even at $\rho=0.1$. The interpretation is as the number of
pairs contributing to each lag bin grows, the per-lag noise floor drops, the physical
peak emerges above it, and the global maximum of the estimated CCF converges onto the
true lag $\tau_0$.
\subsubsection{Peak amplitude}
A systematic, $\rho$-dependent bias is observed: \nufftcf (both kernels) and the
real-space Gaussian estimator all \emph{overestimate} the peak amplitude compared to the
true value, with an absolute bias that decreases with increasing $\rho$: e.g.,
$+0.055$/$+0.058$/$+0.056$ at $\rho=0.2$ versus $+0.012$/$+0.023$/$+0.007$ at
$\rho=0.95$, for \nufftcf Gaussian and rectangle kernels, and real-space Gaussian
respectively. The fact that the real-space evaluation shares essentially the same bias
as the \nufftcf NUFFT-based counterpart shows that this overestimation is inherent to
the kernel-weighted estimator itself, rather than a NUFFT-specific artifact. We discuss
this in more detail below. Note that the rectangle kernel consistently produces a
slightly higher peak amplitude than the Gaussian kernel estimator, as the latter smooths
more strongly over the peak neighborhood.

In contrast, \pyzdcf strongly underestimates the amplitude across the entire tested
range (at best estimating $0.154$ instead of $0.95$), with the absolute gap widening as
$\rho$ increases. This underestimation is again a direct consequence of slotting, as
discussed for the peak location.
\subsubsection{Systematic overestimation by the kernel estimators}
\label{sec:overestimation_peak_ampl}
The real-space Gaussian-kernel and the \texttt{nufftcf} Gaussian-kernel results agree
quite well at every $\rho$ (e.g. $0.538$ vs. $0.539$ at $\rho=0.5$ and $0.957$ vs.
$0.962$ at $\rho=0.95$), sharing essentially the same positive bias. Since the
real-space Gaussian estimator (Equation~\ref{eq:kernel_estimator}) and the \nufftcf
NUFFT-based estimator (Algorithm~\ref{alg:ccf-nufft}, step 10) numerator and denominator
are weighted by the same kernel at the level of individual pairs in both
implementations, the overestimation cannot be attributed to the NUFFT gridding algorithm
(Section~\ref{sec:nufft}). It is instead an intrinsic property of the kernel-weighted
correlation estimator itself.

We attribute this to a \emph{selection} bias: the reported peak is a maximum, over a
dense grid of lags, of a locally noisy estimate -- at each lag only the pairs within a
few kernel bandwidths ($\mathrm{bin\_width}=0.5$~d) contribute, given a mean sampling
interval of $\sim1.7$--$2.5$~d. Scanning many such noisy per-lag estimates and reporting
the global maximum systematically overestimates the true peak, an effect that is largest
when $\rho$ is comparable to the per-lag noise level and vanishes as $\rho$ tends to
$1$, since the Pearson correlation is bounded by unity. This also explains the
$\rho=0.1$ case (excluded from the quantitative analysis below), where the true bump at
$\tau_0$ is not statistically distinguishable from the noise floor and the reported
maximum is essentially a pure noise fluctuation, located at $\mathrm{lag}=167$ rather
than $60$.
\begin{table*}
	\centering
	\caption{Peak bias $\Delta=\mathrm{peak\_val}-\rho$ compared to the
		background noise level $\mathrm{std\_far}$ (std of the CCF estimate over
		$|\mathrm{lag}-\tau_0|>50$~d, $n_\mathrm{far}=79$), for $\rho\ge0.2$.}
	\label{tab:bias_vs_background}
	\resizebox{\textwidth}{!}{%
		\begin{tabular}{l ccc ccc ccc ccc ccc}
			\toprule
			& \multicolumn{3}{c}{$\rho=0.2$} & \multicolumn{3}{c}{$\rho=0.3$} & \multicolumn{3}{c}{$\rho=0.5$} & \multicolumn{3}{c}{$\rho=0.7$} & \multicolumn{3}{c}{$\rho=0.95$} \\
			\cmidrule(lr){2-4}\cmidrule(lr){5-7}\cmidrule(lr){8-10}\cmidrule(lr){11-13}\cmidrule(lr){14-16}
			Method & $\Delta$ & $\mathrm{std\_far}$ & ratio & $\Delta$ & $\mathrm{std\_far}$ & ratio & $\Delta$ & $\mathrm{std\_far}$ & ratio & $\Delta$ & $\mathrm{std\_far}$ & ratio & $\Delta$ & $\mathrm{std\_far}$ & ratio \\
			\midrule
			nufftcf Gaussian   & 0.055 & 0.047 & 1.16 & 0.051 & 0.045 & 1.14 & 0.039 & 0.041 & 0.96 & 0.024 & 0.038 & 0.64 & 0.012 & 0.042 & 0.28 \\
			nufftcf rectangle  & 0.058 & 0.048 & 1.20 & 0.055 & 0.046 & 1.21 & 0.046 & 0.042 & 1.10 & 0.033 & 0.039 & 0.85 & 0.023 & 0.043 & 0.54 \\
			realspace Gaussian & 0.056 & 0.051 & 1.10 & 0.052 & 0.048 & 1.07 & 0.038 & 0.044 & 0.88 & 0.022 & 0.040 & 0.55 & 0.007 & 0.044 & 0.17 \\
			\bottomrule
	\end{tabular}}
\end{table*}

To test this hypothesis quantitatively, for $\rho\ge0.2$ and each estimator we compare
on Table~\ref{tab:bias_vs_background} the peak bias
$$\Delta=\mathrm{peak\ val.}-\rho$$ 
to $\mathrm{std\_far}$ the standard deviation of the CCF estimate away from the peak
defined as
$$
\mathrm{std\_far} = \mathrm{std}\Bigl( CCF(\mathrm{lag}) \mid |\mathrm{lag} - \tau_0| > 5\alpha = 50~\mathrm{d} \Bigr)
$$
The lag window is chosen so that the theoretical CCF (Equation~\ref{eq:ou_theory_ccf})
has decayed to $e^{-5}\approx0.7\%$ of its peak value there. With
$\mathrm{lags}\in[1,180]$~d this leaves $n_\mathrm{far}=79$ lag points. Two features
support the selection bias interpretation.

First, $\mathrm{std\_far}$ is essentially constant across $\rho$
($\approx0.038$--$0.051$) for all three estimators, consistent with a signal-independent
noise floor set by sampling and kernel bandwidth.

Second, $\Delta/\mathrm{std\_far}$ is of order unity at weak-to-moderate coupling
($\approx0.9$--$1.2$ at $\rho=0.2$--$0.5$) -- the peak bias is comparable to the
background noise, exactly as expected when selecting the maximum of a noisy curve -- and
decreases monotonically as $\rho$ grows, down to $\approx0.17$--$0.54$ at $\rho=0.95$,
since $\mathrm{std\_far}$ stays roughly constant while $\Delta$ itself shrinks toward
the $\rho=1$ bound. The rectangle kernel shows the largest bias-to-noise ratio at every
$\rho$, consistent with its sharper effective bandwidth producing a noisier per-lag
estimate, while real-space and \texttt{nufftcf} Gaussian remain statistically
indistinguishable, confirming a kernel-choice effect rather than an implementation
artifact.

This sample-size dependence can also be verified directly. Repeating the experiment with
a ten-fold longer baseline ($36\,500$ instead of $3650$~d) markedly reduces $|\Delta|$
at every tested $\rho$, for all three estimators:
e.g.\ $\Delta=-0.030$/$-0.027$/$-0.029$ at $\rho=0.2$ (versus $+0.055$/$+0.058$/$+0.056$
in the fiducial run) and $-0.008$/$-0.001$/$-0.003$ at $\rho=0.95$ (versus
$+0.012$/$+0.023$/$+0.007$), confirming that the bias shrinks as the per-lag noise floor
drops with increasing sample size. Notably, the residual at this larger sample size
turns slightly \emph{negative} for every estimator, and is consistently less negative
for the rectangle kernel than for the Gaussian kernel -- the same ordering seen in the
selection-bias-dominated regime -- suggesting that once the selection bias is
suppressed, a small residual smoothing bias from averaging the sharply peaked CCF over a
finite kernel bandwidth becomes visible instead.

A natural mitigation, in both implementations (real space or NUFFT-based \nufftcf
computatiions), would be to widen the kernel bandwidth reducing per-lag variance at the
cost of resolution or to calibrate and subtract the selection bias via a
bootstrap/Monte-Carlo procedure on surrogate data, analogous to the internal Monte-Carlo
simulations used by \pyzdcf for its own uncertainty estimates.
Section~\ref{sec:astro_mc} revisits this same diagnosis -- a single-realization
deviation mistaken for a bias -- on an independently-motivated, physical use case, using
repeated realizations rather than a longer baseline to demonstrate it.
%
\section{A physical use-case: Simulated Astrophysical Light Curve}
\label{sec:astro_example}
The synthetic series of Section~\ref{sec:comparison} were designed to isolate ACF/CCF
estimator behaviour under simple, well-controlled statistical models, so as to enable a
quantitative, independent cross-check against \pastas and \pyzdcf. To illustrate
\nufftcf on a case closer to observational practice, we simulate a single stellar
photometric light curve combining a quasi-periodic rotation signal, correlated
``flicker'' noise, realistic ground-based sampling gaps, and heteroscedastic measurement
noise, and use \nufftcf alone to recover the injected rotation period and noise
correlation time.\footnote{See \texttt{nufftcf\_astro\_demo.ipynb} notebook.}
\subsection{Simulated light curve}
\label{sec:astro_model}
The latent flux is
\begin{equation}
y(t) = A(t)\cos\!\left(\frac{2\pi t}{P_{\rm rot}}+\varphi\right) + n(t),
\label{eq:astro_model}
\end{equation}
where $\varphi$ is a random phase, uniform on $[0,2\pi)$ and independent of $A(t)$;
$A(t)$ is a stationary Ornstein--Uhlenbeck (OU) process of mean $A_0$, variance
$\sigma_A^2$, and correlation time $\tau_A \gg P_{\rm rot}$, modelling slow spot
evolution; and $n(t)$ is an independent OU process of variance $\sigma_n^2$ and
correlation time $\tau_n$, modelling short-timescale correlated (flicker/granulation)
noise. The normalized autocorrelation of $y(t)$ is then analytic,
\begin{equation}
\rho_{\rm true}(\tau) = \frac{\tfrac12\cos\!\left(\frac{2\pi\tau}{P_{\rm rot}}\right)\left(\sigma_A^2 e^{-|\tau|/\tau_A}+A_0^2\right) + \sigma_n^2 e^{-|\tau|/\tau_n}}{\tfrac12(\sigma_A^2+A_0^2)+\sigma_n^2},
\label{eq:astro_acf_true}
\end{equation}
and has the form of a sum of an exponentially damped cosine and an exponential decay,
closely related to the stochastically driven damped harmonic oscillator (SHO) kernels of
the celerite family \citep{ForemanMackey2017}, which have been used to model stellar
rotation \citep[e.g.][]{Angus2018} and granulation/flicker noise in a similar
decomposition. It is qualitatively similar in spirit, though not identical in functional
form, to the product-form quasi-periodic kernels (squared-exponential times periodic)
commonly used in Gaussian-process models of stellar activity
\citep[e.g.][]{Rajpaul2015}. We use $P_{\rm rot}=12$~d, $A_0=1$, $\sigma_A=0.35$,
$\tau_A=220$~d, $\sigma_n=0.55$, $\tau_n=3.5$~d. Both $A(t)$ and $n(t)$ are simulated
directly at the (irregular) observation epochs using the exact OU transition kernel, so
no dense-grid discretization is needed.

\begin{figure}
\centering
\includegraphics[width=\columnwidth]{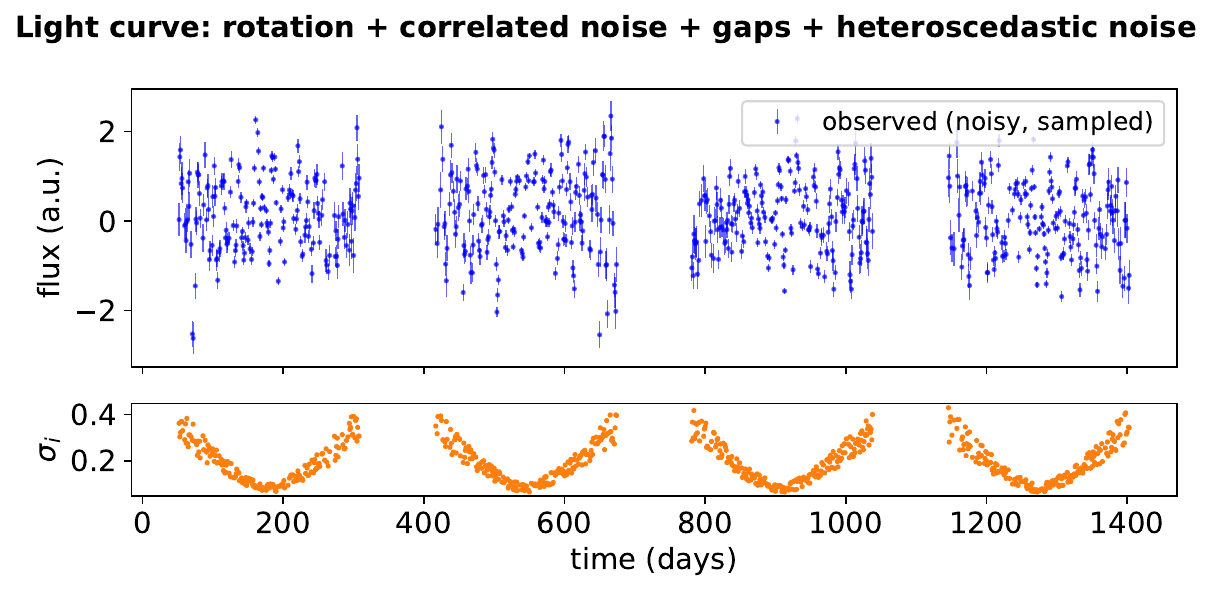}
\caption{Simulated ground-based light curve (top; error bars show
$\sigma_i$) combining stellar-rotation modulation, correlated OU
noise, seasonal/weather gaps, and heteroscedastic measurement noise
(bottom).}
\label{fig:astro_lightcurve}
\end{figure}

Ground-based sampling is emulated over four years with one nightly epoch, an
$\sim$8-month seasonal observability window, and $25\%$ of otherwise-observable nights
lost to weather, yielding a fiducial realization of $n=762$ irregular epochs (median
cadence $1.0$~d within a season). Each point additionally carries a heteroscedastic
measurement uncertainty $\sigma_i$, growing toward the edges of the observing season
(higher airmass), with $\sigma_i \in [0.07, 0.43]$~d (median $0.18$) for this parameter
choice; the observed flux is $x_{\rm obs}(t_i)=y(t_i)+\varepsilon_i$,
$\varepsilon_i\sim\mathcal N(0,\sigma_i^2)$. Figure~\ref{fig:astro_lightcurve} shows the
resulting light curve.
\subsection{ACF recovery}
\label{sec:astro_acf}
\begin{figure}
\centering
\includegraphics[width=\columnwidth]{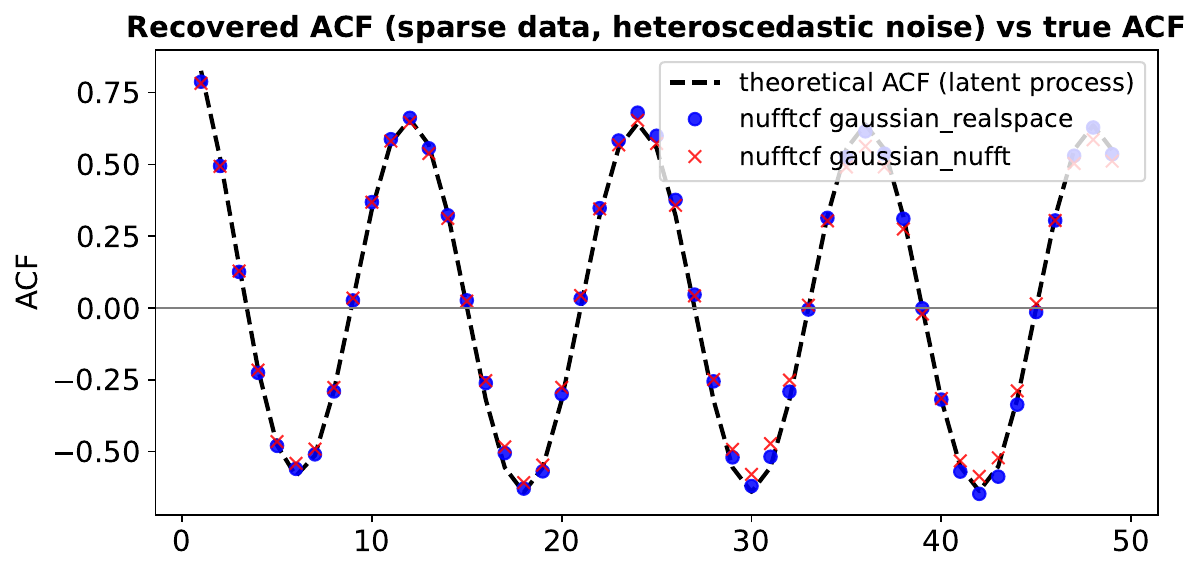}
\caption{ACF recovered by \nufftcf Gaussian real-space and NUFFT
estimators from the sparse, heteroscedastic light curve of
Figure~\ref{fig:astro_lightcurve}, against the theoretical
latent-process ACF (Equation~\ref{eq:astro_acf_true}).}
\label{fig:astro_acf}
\end{figure}

We compute the ACF with \nufftcf Gaussian-kernel real-space and NUFFT estimators
($\mathrm{bin\_width}=0.5$~d) on $x_{\rm obs}$, at every integer lag from 1 to 49~d.
Since \nufftcf standardizes the series before evaluating $\hat\rho$, and the measurement
noise only contributes at lag-0 (known as ``nugget'' effect \citep[e.g.][]{Sayer2020} or
``zero-lag'' spike \citep[e.g.][]{Lacki:2025yrx}), the observed ACF is a rescaled
version of Equation~\ref{eq:astro_acf_true}:
$\rho_{\rm obs}(\tau) = \kappa\,\rho_{\rm true}(\tau)$, with dilution factor
\begin{align}
\kappa &= \mathrm{Cov}_{\rm tot}(0)/\mathrm{Var}(x_{\rm obs}) \nonumber \\
&= 
\mathrm{Cov}_{\rm tot}(0)/(\mathrm{Cov}_{\rm tot}(0)+\langle\sigma_i^2\rangle) < 1.
\end{align}
Figure~\ref{fig:astro_acf} compares the two \nufftcf estimates to $\rho_{\rm true}$ for
the fiducial realization. Both agree closely with the theoretical curve and with each
other: RMSE $=0.027$ for real-space, $0.040$ for NUFFT, against $\rho_{\rm true}$,
i.e.\ without correcting for the dilution factor above. The rotation-driven oscillation
and the fast initial decay from the flicker noise are both clearly resolved despite the
seasonal gaps and the heteroscedastic noise.
\subsection{Recovering the rotation period and noise timescale}
\label{sec:astro_fit}
We extract two physical quantities from the recovered ACF: the rotation period
$P_{\rm rot}$, taken as the lag of the ACF maximum in a search window around the
injected value, and the flicker-noise correlation time $\tau_n$, obtained by fitting the
short-lag ($\le 14$~d) ACF to a two-component model,
\begin{equation}
\hat\rho(\tau) = \kappa\left[w\cos\!\left(\frac{2\pi\tau}{P_{\rm rot}}\right) + (1-w)\,e^{-\tau/\tau_n}\right],
\label{eq:astro_fit_model}
\end{equation}
with free parameters $(w, \tau_n, P_{\rm rot}, \kappa)$. Note that a naive
single-exponential decay does not fit this range, since $P_{\rm rot}=12$~d is not slowly
varying compared to the fitted lags. The explicit dilution factor $\kappa$ is required
in Equation~\ref{eq:astro_fit_model}: without it, the fit compensates for the
noise-induced amplitude deficit at $\tau>0$ by biasing $\tau_n$ low, since $\hat\rho$
would otherwise be pinned to $1$ at $\tau=0$ by construction.

For the fiducial realization, the ACF peak search recovers $P_{\rm rot}=12.00$~d exactly
(injected: $12.00$~d) for both estimators. The composite fit gives
$\tau_n=4.09\pm0.38$~d (real-space) and $4.16\pm0.38$~d (NUFFT), compared to the
injected $\tau_n=3.5$~d, and $\kappa=0.938\pm0.019$ and $0.940\pm0.020$ respectively, in
agreement with an independent, fit-free estimate
$\kappa_{\rm est}=1-\langle\sigma_i^2\rangle/\mathrm{Var}(x_{\rm obs})=0.954$ obtained
directly from the data.
\subsection{Monte-Carlo over independent realizations}
\label{sec:astro_mc}
\begin{figure}
\centering
\includegraphics[width=\columnwidth]{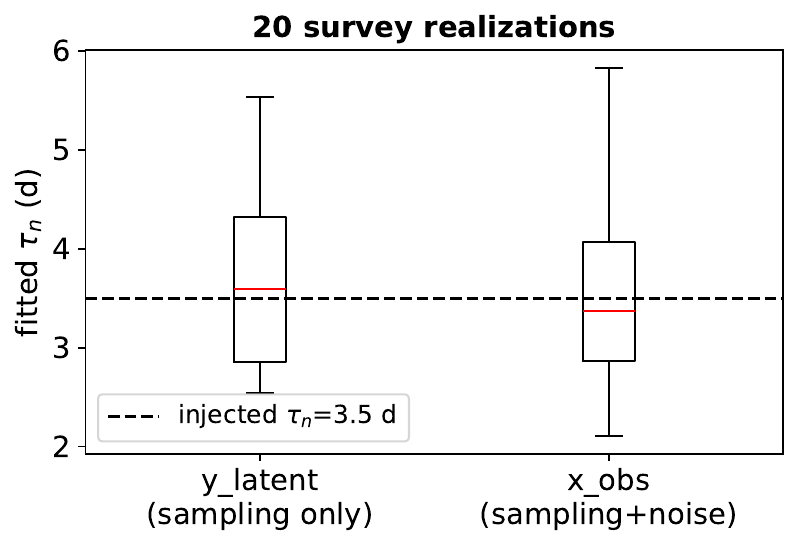}
\caption{Fitted $\tau_n$ over $n_{\rm mc}=20$ independent survey
realizations, for the sampled-only latent signal and the
sampled$+$noisy observed series, against the injected value (dashed).
Medians are shown as red lines, box edges represent the first (Q1) and third (Q3) quartiles, and whiskers extend to the most extreme data point within 1.5 times the interquartile range (IQR) from the boxes.}
\label{fig:astro_mc}
\end{figure}

Taken at face value, the single-realization $\tau_n$ fit above ($\sim4.1$~d) is above
the injected value ($3.5$~d) by more than its formal uncertainty, suggestive of a
residual bias -- reminiscent of the CCF peak-amplitude offsets identified from a single
realization in Section~\ref{sec:overestimation_peak_ampl}. Following the same logic, we
test whether this is a genuine estimator bias or a finite-sample fluctuation by
repeating the full pipeline -- survey sampling, heteroscedastic noise, latent signal,
\nufftcf Gaussian-NUFFT ACF, and the fit of Equation~\ref{eq:astro_fit_model} -- over
$n_{\rm mc}=20$ independent realizations (Figure~\ref{fig:astro_mc}). The same model is
fit to the noiseless theoretical curve, to the noise-free but irregularly-sampled latent
signal, and to the fully observed (sampled$+$noisy) series.

Averaged over the 20 realizations, the fit recovers $\tau_n = 3.64\pm0.19$~d (sampling
only) and $3.60\pm0.24$~d (sampling$+$noise) [mean $\pm$ SEM\footnote{SEM stands for
Standard Error of the Mean.}], both consistent with the injected $\tau_n=3.5$~d, while
the noiseless theoretical curve, which carries no randomness and is therefore fit only
once, returns $\tau_n=3.46$~d. The per-realization scatter is substantial
($\mathrm{std}\simeq0.86$--$1.09$~d), comparable to, or larger than, the $\sim0.6$~d
offset seen for the single fiducial realization: the latter is not statistically
distinguishable from the sampling noise floor. As in the CCF peak-amplitude case of
Section~\ref{sec:overestimation_peak_ampl}, an apparent bias identified from a single,
or a small number of, realization(s) is again a finite-sample/selection effect rather
than an intrinsic property of the \nufftcf estimator. This bias shrinks on average when
several independent realizations are combined, even though the per-realization scatter
remains unchanged, as it is determined by the survey information content rather than by
the estimator itself.

This test reinforces, on an independent, physically-motivated case, the guidance of
Section~\ref{sec:overestimation_peak_ampl}: a single-realization deviation from the
truth should always be weighed against the estimator sampling variance (accessible here
via repeated simulation, or in practice via bootstrap on the real data) before being
interpreted as a bias.
%
\section{Timing Benchmarks}
\label{sec:benchmarks}
To quantify the practical speed-up offered by \nufftcf NUFFT, we benchmarked \nufftcf
and \pastas on synthetic white-noise series of increasing length, separately for
irregularly-sampled and regularly-sampled data. \pyzdcf was excluded from this
comparison: its $O(n^2)$ real-space correlation, combined with a Monte-Carlo error
estimate, makes it prohibitively slow for series of the lengths considered here --
unsurprising, since \pyzdcf was designed to prioritize error estimation over
computational speed.

Benchmarks are restricted to ACF, rather than CCF, for two reasons: first, this allows a
direct comparison against \pastas, which only implements ACF; second, within \nufftcf
itself, CCF and ACF share the same computational cost, since both are driven by the same
NUFFT calls and the same $b$-factor (pair-count normalization) computation.

All benchmarks sweep the series length $n$ at a fixed number of requested lags, $K=366$
(integer lags $0,1,\dots,365$~d), which sits moderately above the measured Gaussian
crossover $K^*$ and well below the rectangle one; the dependence on $K$, which matters
when comparing the NUFFT and real-space estimators, is studied separately in
Section~\ref{sec:discussion_choice}.

For each configuration, computation time was measured over several repeats per series
length, with execution order randomized across (kernel, length) combinations to avoid
confounding a systematic drift (e.g.\ thermal throttling over a long benchmark run) with
the trend in series length being measured. The reported point estimate is the per-length
median (regular case) or minimum (irregular case) computation time, and error bars show
the min-to-max spread across repeats. For each (kernel, algorithm) pair, timings were
fit to the scaling models of Section~\ref{sec:bench_model} using a robust nonlinear
least-squares fit (\texttt{scipy.optimize.least\_squares} \cite{scipy2020},
\texttt{soft\_l1} loss), which down-weights the influence of occasional outlier
measurements without requiring them to be manually identified and removed. Parameter
uncertainties were obtained by bootstrapping over the recorded repeats (500 iterations):
at each iteration, repeats at every series length are resampled with replacement, the
point estimate is recomputed, and the model is refit; the standard deviation of the
refit parameters across iterations gives the reported uncertainty.

For the sake of completeness and reproducibility, we detail the exact hardware
configuration used to conduct the benchmarks: this is a MacBook Pro (2020) equipped with
a 2 GHz Intel Core i5 processor (quad-core), 512 KB of L2 cache per core, and 6 MB of
shared L3 cache, along with 16 GB of LPDDR4X RAM (3733 MHz, dual-channel, 8 GB per
channel), running macOS 26.5.2 (build 25F84).
\subsection{Estimator families and cost model}
\label{sec:bench_model}
Table~\ref{tab:families} groups the ACF and CCF estimators of \nufftcf into their three
families and gives their cost for $n$ samples and $K$ requested lags. The NUFFT
estimators combine two ingredients of a different nature
(Section~\ref{sec:nufftcf_approach}): the FINUFFT calls, of cost $O(n\log n)$ (as
$N_1\propto n$) plus $O(K)$, and the normalization by the effective pair count, an
$O(n)$ two-pointer scan \emph{per lag}, hence $O(nK)$. We therefore fit their timings,
at fixed $K$, with the two-term model
\begin{equation}
	t_{\rm nufft}(n) = c_n\, n + c_{n\ln n}\, n\ln n + t_0,
	\label{eq:cost_nufft}
\end{equation}
where $c_n\propto K$ carries the pair-count scan, $c_{n\ln n}$ the FINUFFT calls, and
$t_0$ is a fixed overhead.

The other timings are fit with a single $n$-dependant term:
\begin{itemize}
\item $c_{n^2}\,n^2 + t_0$ \inlineeqnum{eq:model1} for the \pastas Gaussian/rectangle
bin methods,
\item $c_n\,n + t_0$ \inlineeqnum{eq:model2} for its \texttt{"regular"} bin method and
the real-space estimators,
\item $c_{n\ln n}\,n\ln n + t_0$ \inlineeqnum{eq:model3} for the \nufftcf FFT path,
whose cost does not depend on $K$.
\end{itemize}
At fixed $K$, the two terms of Equation~\ref{eq:cost_nufft} are hardly distinguishable
over one or two decades in $n$, and $n\ln n$ overtakes $nK$ only for $n$ of order
$e^{K}$, out of reach for any realistic $K$. What the fits can tell is therefore which
term dominates the measured time, rather than a sharp discrimination of the exponent. It
is the pair-count term for the Gaussian kernel, whose inner loop evaluates an
exponential for every point in the kernel window, and a comparable share of both terms
for the rectangle kernel, whose inner loop reduces to pointer arithmetic (see below).
\begin{table*}
		\centering
		\small
		\renewcommand{\arraystretch}{1.15}
		\setlength{\tabcolsep}{4pt}
		\caption{Estimator families of \nufftcf (ACF and CCF merged) and their cost for $n$ samples and $K$ requested lags.
			Functions are named \texttt{compute\_[acf|ccf]\_[gaussian|rectangle]\_[nufft|realspace|fft]} and share the calling
			convention \texttt{fn(lags, t, x, [s, y], bin\_width)}$\,\to(c,b)$; \texttt{compute\_acf\_regular\_fft} has no CCF counterpart.
			$K^*$ is the number of lags above which the NUFFT estimator is faster than the real-space one (Section~\ref{sec:discussion_choice}).}
		\label{tab:families}
		\begin{tabular}{@{}>{\raggedright\arraybackslash}p{0.17\textwidth}>{\raggedright\arraybackslash}p{0.17\textwidth}>{\raggedright\arraybackslash}p{0.19\textwidth}>{\raggedright\arraybackslash}p{0.35\textwidth}@{}}
			\toprule
			Family & Sampling & Cost & Method and remarks \\
			\midrule
			\texttt{\_nufft} (Gaussian, rectangle) & irregular &
			$O(n\log n)+O(nK)\;\Rightarrow\;\sim O(nK)$ &
			NUFFT + Wiener--Khinchin for the numerator, one two-pointer scan for the denominator. Faster than \texttt{\_realspace} for $K>K^*$, with $K^*\simeq120$--$280$ (Gaussian, no clear trend with $n$ over $n=10^3$--$10^5$) and $K^*\gtrsim2000$ (rectangle, growing with $n$). $\sim$1--3\% residual bias on strongly periodic, gappy signals. \\
			\texttt{\_realspace} (Gaussian, rectangle) & irregular or regular &
			$O(nK)$ &
			Direct kernel-weighted sum, two two-pointer scans (numerator and denominator). Exact reference, no periodicity assumption; competitive or faster than \texttt{\_nufft} for $K<K^*$. \\
			\texttt{\_fft} (Gaussian, rectangle) & regular only (CCF: common $\Delta t$ and integer lattice) &
			$O(n\log n)$, independent of $K$ &
			Classic FFT correlation and discrete smoothing filter; the pair count comes from filtering the deterministic raw-pair ramp, without any scan. Fastest on regular data. \\
			\texttt{acf\_regular\_fft} (ACF only, no kernel) & regular only &
			$O(n\log n)$, independent of $K$ &
			Reproduces the \pastas \texttt{"regular"} bin method to numerical precision. \\
			\bottomrule
		\end{tabular}
\end{table*}
\subsection{Irregularly-sampled data}
\label{sec:bench_irregular}
\begin{figure*}
	\centering
	\includegraphics[width=0.85\textwidth]{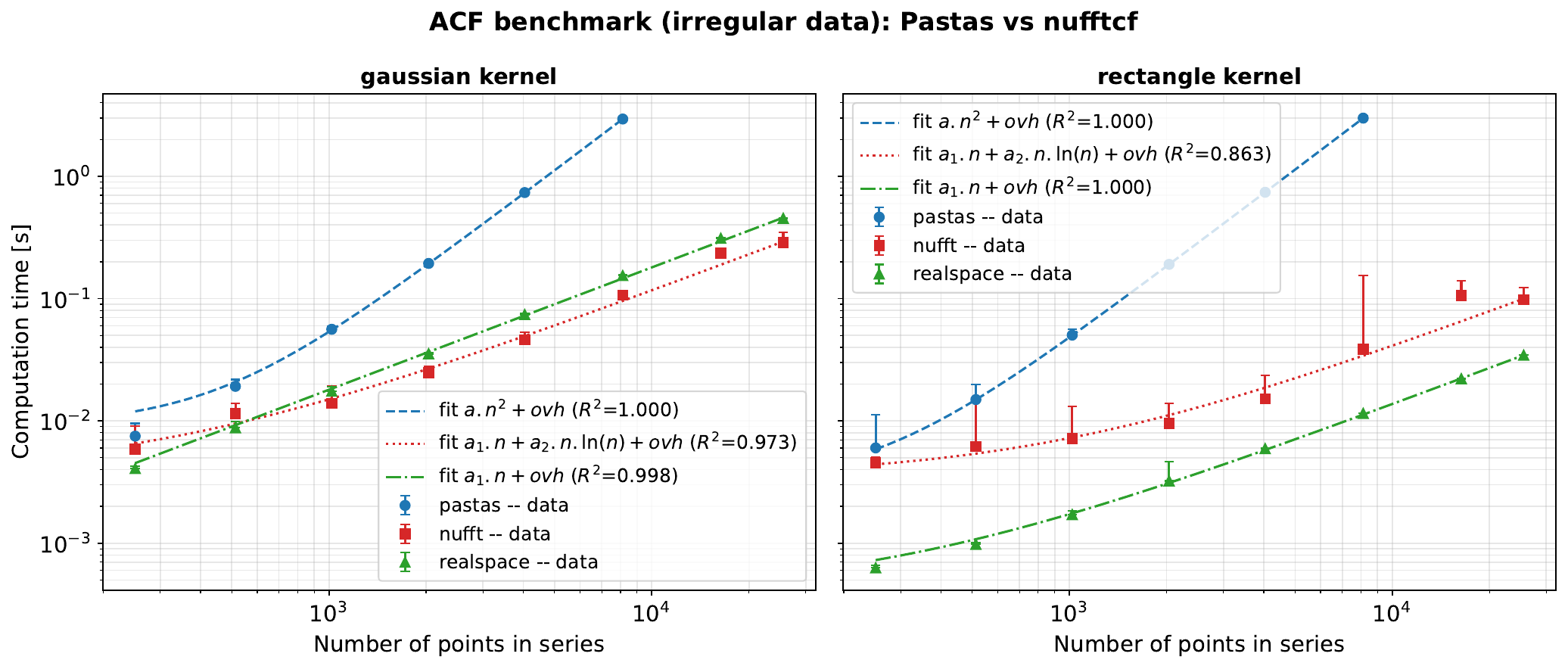}
	\caption{ACF benchmark, \nufftcf versus \pastas, on irregularly-sampled
		data (Gaussian and rectangle kernels): fit of \pastas (dashed), of the \nufftcf NUFFT
		estimator (dotted) and of the \nufftcf real-space estimator (dash-dotted).
		Error bars show the min-to-max spread across repeats at each point. The number of lags is $K=366$ which
		is well moderately above the Gaussian crossover $K^\ast$ and well below the rectangle
		one (Section~\ref{sec:discussion_choice}).}
	\label{fig:acf_bench}
\end{figure*}

Irregularly-sampled series are the primary use case \nufftcf was designed for, and the
NUFFT-based and real-space estimators were accordingly compared against \pastas for both
the Gaussian and rectangle smoothing kernels. Table~\ref{tab:fit_results} summarizes the
fitted parameters, and Figure~\ref{fig:acf_bench} shows the results.

As expected from the underlying algorithms, \pastas slotting technique scales
 quadratically in the number of points. The \nufftcf NUFFT timings are fit with the
 two-term model of Equation~\ref{eq:cost_nufft}, and the real-space timings with the
 single-term model of Equation~\ref{eq:model2}, the same functional form as the \pastas
 \texttt{"regular"} bin method.

For both kernels the NUFFT fit drives the $n\ln n$ coefficient to its lower bound:
$c_{n\ln n}$ is compatible with zero, and the time is carried by the term linear in $n$,
$c_n=(1.13\pm0.09)\times10^{-5}$~s (Gaussian) and $(3.77\pm0.44)\times10^{-6}$~s
(rectangle), i.e. by the $O(n)$-per-lag pair-count scan rather than by the FINUFFT
calls. For $K=366$ this amounts to $\simeq31$ and $\simeq10$~ns per point and per lag,
respectively. The rectangle scan is $\sim3.0\times$ cheaper per point, as expected since
it needs no exponential evaluation\footnote{The rectangle NUFFT fit is less constrained,
$R^2=0.86$, because of the larger scatter of its repeated measurements.}. With $n$
spanning only about two decades at a fixed $K$, the $n$ and $n\ln n$ terms are nearly
degenerate, so the meaningful result is which term dominates rather than a sharp
discrimination of the exponent.

The real-space fits are tightly constrained ($R^2=1.00$ for both kernels) and give
$c_n=(1.80\pm0.00)\times10^{-5}$~s (Gaussian) and $(1.34\pm0.00)\times10^{-6}$~s
(rectangle), with a negligible overhead ($t_0\lesssim0.4$~ms). At the fixed $K=366$ used
throughout this section, the real-space estimator is slower than NUFFT for the Gaussian
kernel and faster for the rectangle kernel which is consistent with $K$ lying above the
Gaussian crossover $K^*$ and inside the rectangle one, as discussed in
Section~\ref{sec:discussion_choice}.

The quadratic-versus-linear (at fixed $K$) scaling translates into a rapidly growing
speed advantage for \nufftcf. At $n=8000$, the largest length at which \pastas was
measured, the fitted times are $2.85$~s versus $0.094$~s for the Gaussian kernel
($\sim30\times$) and $2.90$~s versus $0.034$~s for the rectangle kernel
($\sim86\times$). The real-space estimator shows the same advantage over \pastas:
$0.14$~s for the Gaussian kernel ($\sim20\times$) and $0.011$~s for the rectangle kernel
($\sim260\times$). Thanks to a small overhead ($\lesssim4$~ms for NUFFT,
$\lesssim0.4$~ms for real-space) \nufftcf is advantageous compared to \pastas already
from a few hundred points, well before the asymptotic regime is reached.

\begin{table*}
\centering
\small
\setlength{\tabcolsep}{7pt}
\caption{Fit results for irregularly- and regularly-sampled data, at a fixed number of requested lags $K=366$ (coefficients in seconds, overhead $t_0$ in ms; uncertainties are bootstrap standard deviations). Models are described in Section~\ref{sec:bench_model} (Equations~\ref{eq:cost_nufft}-\ref{eq:model3}). For the NUFFT path the $n\ln n$ coefficient sits at its lower bound and is compatible with zero, except for the rectangle kernel on regularly-sampled data.}
\label{tab:fit_results}
\resizebox{\textwidth}{!}{%
\begin{tabular}{>{}l>{}l>{}c>{}c>{}c>{}r>{}r}
\toprule
Kernel & Algorithm & $c_{n^2}$ & $c_{n}$ & $c_{n\ln n}$ & $t_0$ [ms] & $R^2$ \\
\midrule
\multicolumn{7}{>{}l}{\emph{Irregular sampling}} \\
Gaussian & \pastas & $(4.44\pm0.01)\times10^{-8}$ & -- & -- & $9.17\pm0.26$ & 1.00 \\
Gaussian & \nufftcf (nufft) & -- & $(1.13\pm0.09)\times10^{-5}$ & $(0.00\pm1.31)\times10^{-7}$ & $3.69\pm0.53$ & 0.97 \\
Gaussian & \nufftcf (realspace) & -- & $(1.80\pm0.00)\times10^{-5}$ & -- & $0.00\pm0.15$ & 1.00 \\
rectangle & \pastas & $(4.52\pm0.02)\times10^{-8}$ & -- & -- & $3.01\pm0.65$ & 1.00 \\
rectangle & \nufftcf (nufft) & -- & $(3.77\pm0.44)\times10^{-6}$ & $(0.00\pm0.62)\times10^{-7}$ & $3.46\pm0.46$ & 0.86 \\
rectangle & \nufftcf (realspace) & -- & $(1.34\pm0.00)\times10^{-6}$ & -- & $0.39\pm0.03$ & 1.00 \\
\midrule
\multicolumn{7}{>{}l}{\emph{Regular sampling}} \\
Gaussian & \pastas & $(4.79\pm0.02)\times10^{-8}$ & -- & -- & $4.40\pm1.30$ & 1.00 \\
Gaussian & \nufftcf (fft) & -- & -- & $(2.18\pm0.19)\times10^{-8}$ & $0.78\pm0.21$ & 0.98 \\
Gaussian & \nufftcf (nufft) & -- & $(1.65\pm0.03)\times10^{-5}$ & $(0.00\pm2.70)\times10^{-8}$ & $4.58\pm0.89$ & 1.00 \\
rectangle & \pastas & $(4.63\pm0.02)\times10^{-8}$ & -- & -- & $4.90\pm1.30$ & 1.00 \\
rectangle & \nufftcf (fft) & -- & -- & $(1.26\pm0.02)\times10^{-8}$ & $0.69\pm0.19$ & 0.89 \\
rectangle & \nufftcf (nufft) & -- & $(2.18\pm0.23)\times10^{-6}$ & $(2.12\pm0.17)\times10^{-7}$ & $6.69\pm0.19$ & 0.98 \\
regular & \pastas & -- & $(1.39\pm0.02)\times10^{-5}$ & -- & $18.72\pm0.67$ & 1.00 \\
regular & \nufftcf (fft) & -- & -- & $(9.50\pm0.10)\times10^{-9}$ & $0.064\pm0.009$ & 0.99 \\
\bottomrule
\end{tabular}}
\end{table*}

\begin{figure*}
	\centering
	\includegraphics[width=0.5\textwidth]{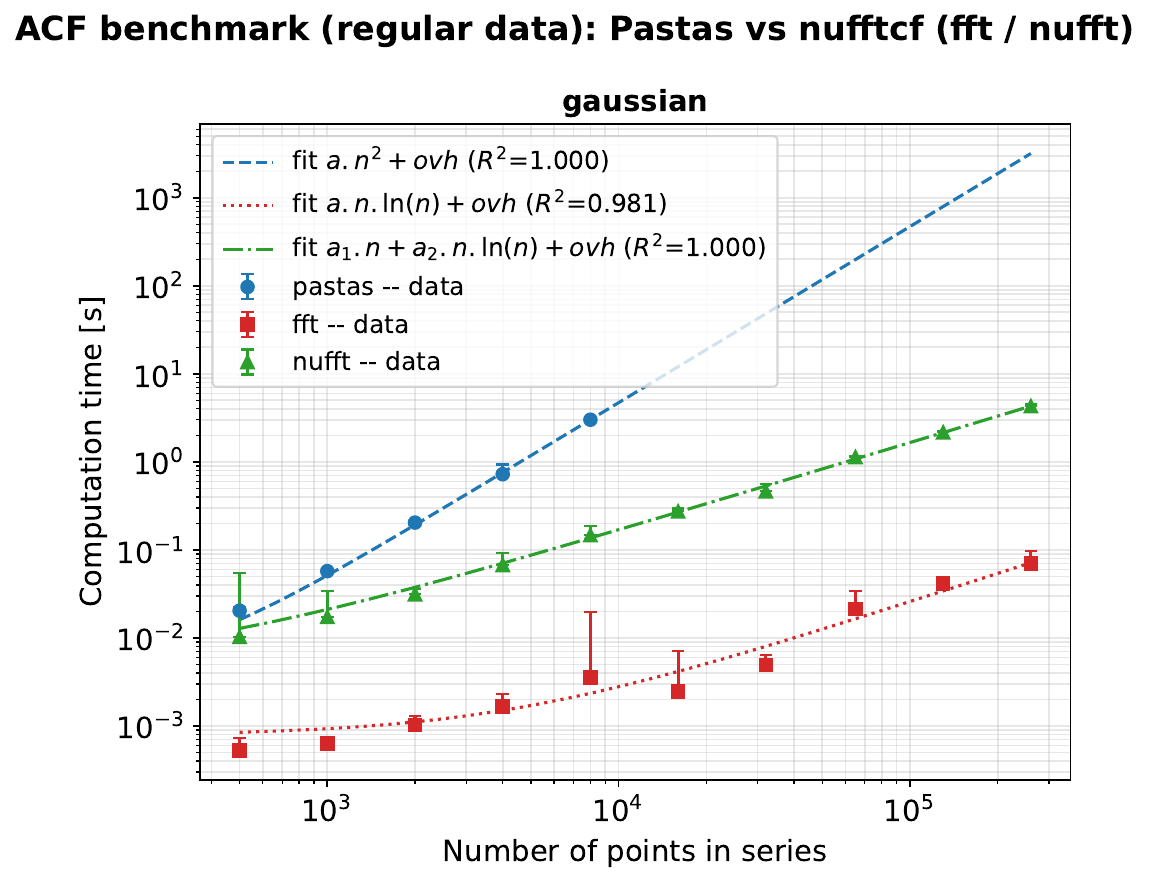}
	\caption{ACF benchmark on regularly-sampled data, \pastas versus
		\nufftcf (dedicated FFT path and general-purpose NUFFT path),
		Gaussian kernel: fit of \pastas (dashed), FFT (dotted) and NUFFT (dash-dotted).}
	\label{fig:acf_regular_bench}
\end{figure*}
\subsection{Regularly-sampled data}
When the data lie on a uniform grid, there is no need to pay for the generality of a
NUFFT: \nufftcf also provides a dedicated classic-FFT path (\texttt{fft\_acf.py}) that
reproduces the exact same Gaussian/rectangle kernel definitions as the NUFFT and
real-space estimators, but via a plain FFT correlation with no numba/finufft dependency
in the hot path. To quantify what this dedicated fast path buys over simply reusing the
general-purpose NUFFT estimator on regular data, we benchmarked \pastas against both the
FFT and NUFFT paths of \nufftcf for the Gaussian and rectangle kernels, and additionally
against \pastas own \texttt{"regular"} bin method (a fixed-size windowed
\texttt{numpy.corrcoef}\footnote{\url{https://numpy.org/doc/stable/reference/generated/numpy.corrcoef.html}}
per lag, with no smoothing kernel), for which the comparison is limited to the dedicated
\texttt{fft\_acf} no-kernel estimator, since NUFFT has no counterpart for this case.
Note that \pastas \texttt{"regular"} bin method scales empirically as $O(n)$, unlike its
$O(n^2)$ \texttt{"Gaussian"}/\texttt{"rectangle"} bin methods.

Fit results are given in Table~\ref{tab:fit_results}, and the Gaussian kernel comparison
is shown in Figure~\ref{fig:acf_regular_bench}. The dedicated FFT path is markedly
faster than the NUFFT path on regular data. At $n=2.56\times10^{5}$ the fitted times are
$0.07$~s versus $4.2$~s for the Gaussian kernel ($\sim60\times$) and $0.04$~s versus
$1.2$~s for the rectangle kernel ($\sim30\times$), and its overhead $t_0$ is
$\sim$6--10$\times$ lower ($0.7$--$0.8$~ms versus $4.6$--$6.7$~ms). This is expected:
the FFT path obtains the effective pair count by filtering the deterministic pair-count
ramp, so its cost is $O(n\log n)$ and independent of $K$, whereas the NUFFT path still
pays the $O(nK)$ two-pointer scan. Here again the Gaussian NUFFT fit is carried by the
linear pair-count term ($c_{n\ln n}$ compatible with zero), while for the rectangle
kernel both terms are resolved and contribute comparably ($\sim45\%/55\%$ of the fitted
time at $n=2.56\times10^5$).

Both \nufftcf paths remain far below the $O(n^2)$ \pastas Gaussian/rectangle bin
methods. The FFT path also beats \pastas own $O(n)$ \texttt{"regular"} bin method by a
factor $\sim100$ at that length, whose much larger leading coefficient outweighs its
more favorable scaling.

The real-space estimator was not separately re-benchmarked here: its cost model
(Equation~\ref{eq:model2}), does not depend on whether the sampling is regular or
irregular, and the comparison of interest on a uniform grid is between the classic-FFT
and NUFFT paths, both of which the dedicated FFT path outperforms.
%
\section{Discussion on \nufftcf}
\label{sec:discussion}
%
\subsection{Choosing an estimator: NUFFT, real-space, or classic-FFT}
\label{sec:discussion_choice}
The three estimator families provided by \nufftcf share a single calling convention
(Section~\ref{sec:nufftcf_approach}) so the choice can be made per use case. For
\emph{regularly-sampled} data, the dedicated classic-FFT (Section~\ref{sec:benchmarks})
should always be preferred: it reproduces the same Gaussian/rectangle kernel definitions
at the lowest cost, with no NUFFT-specific approximation. For \emph{irregularly-sampled}
data---the intended use case for \nufftcf---both the NUFFT and the real-space estimators
cost $O(nK)$ and outperform the $O(n^2)$ slotting (Table~\ref{tab:fit_results}) already
for a few hundred points (Section~\ref{sec:bench_irregular}).
	
The benchmarks of Section~\ref{sec:bench_irregular} sweep $n$ at a fixed $K$ and cannot
separate the NUFFT and real-space estimators, which are both $O(nK)$ and differ only by
their prefactors. Both rely on the same $O(n)$-per-lag two-pointer scan (possible
because $t$ is sorted), but \texttt{\_realspace} runs it twice, for the numerator and
for the denominator, whereas \texttt{\_nufft} runs it once, for the denominator only,
the numerator coming from the NUFFT evaluation. The saved pass must amortize the fixed
cost of the FINUFFT calls, so the faster estimator depends on $K$, not on $n$. For the
Gaussian kernel, $K^*$ is decreasing with $n$ and is of the order of $100$--$300$ for
$n$ in the range of $10^3$--$10^5$. For the rectangle kernel, in the same range of $n$,
the real-space estimator remains faster than NUFFT with $K^\ast\approx 3000$ for
$n\approx 10^5$. At the value of $K=366$ used in the benchmark of
Section~\ref{sec:bench_irregular}, $K$ sits moderately above $K^*$ for the Gaussian
kernel, where NUFFT keeps a modest edge, and well below $K^*$ for the rectangle kernel,
where real-space is markedly faster.
	
Since the Gaussian kernel is the one that performs best, or close to best, in the
benchmark of \cite{Rehfeld2011}, it remains the natural choice, together with the NUFFT
estimator, at typical lag counts; for the rectangle kernel, however, real-space should
be preferred by default at these lag counts. The real-space estimator also remains
preferable, regardless of kernel, when only a handful of lags is needed, when the small
residual bias of the NUFFT estimator on strongly periodic, gappy signals must be avoided
entirely, and as an exact reference against which a NUFFT-based result can be checked on
a subset of the data.
	
Table~\ref{tab:practical_takeaways} summarizes these practical takeaways.
\begin{table*}
	\centering
	\small
	\renewcommand{\arraystretch}{1.25}
	\caption{Practical takeaways for choosing an ACF/CCF estimator with \nufftcf.}
	\label{tab:practical_takeaways}
	\begin{tabular}{@{}>{\raggedright\arraybackslash}p{0.27\textwidth}>{\raggedright\arraybackslash}p{0.63\textwidth}@{}}
		\toprule
		Situation & Recommendation \\
		\midrule
		\nufftcf versus \pastas and \pyzdcf &
		For any realistic $K\ll n$, \nufftcf is far faster than the $O(n^2)$ methods. \\
		NUFFT versus real-space (irregular data) &
		NUFFT keeps a modest edge above $K^*$ for the Gaussian kernel only ($K^*\simeq100$--$380$); for the rectangle
		kernel, real-space is faster over essentially the whole practical range of $K$ ($K^*\gtrsim2000$). Use real-space
		by default for the rectangle kernel, and for a handful of lags or as an exact reference in general. \\
		Regular data &
		The dedicated FFT path, whose cost is independent of $K$, is preferable to both the NUFFT and real-space
		estimators. \\
		Kernel choice &
		The Gaussian kernel is the natural default (best, or close to best, in the benchmark of \cite{Rehfeld2011}). BHowever,
		the rectangle kernel may be valuable for comparaison.\\
		\bottomrule
	\end{tabular}
\end{table*}	
\subsection{Limitations}
\label{sec:discussion_limitations}
Two limitations, quantified in Section~\ref{sec:comparison}, are worth restating as
guidance rather than as a defect specific to \nufftcf. First, the Gaussian- and
rectangle-kernel estimators, whether evaluated in real space or via NUFFT, share an
intrinsic selection bias. This has led to overestimating a CCF peak amplitude by an
amount of order the background noise level of the estimate
(Table~\ref{tab:bias_vs_background}, Figure~\ref{fig:ccf_ou}). This is a property of
maximizing over many noisy per-lag estimates, not a NUFFT-specific artifact
(Section~\ref{sec:overestimation_peak_ampl}), and it shrinks as the sampling density
increases: a ten-fold larger sample already reduces the bias by a factor of a few and
leaves only a small residual smoothing bias from averaging the CCF over a finite kernel
bandwidth $h$.

The same diagnosis extends beyond the CCF peak amplitude itself: in the
physically-motivated light-curve case of Section~\ref{sec:astro_mc}, a
single-realization time-delay fit that appeared offset from the injected value by more
than its formal uncertainty turned out, once averaged over $n_{\rm mc}=20$ independent
realizations, to be statistically consistent with the truth -- confirming that the
apparent bias was a finite-sample selection effect rather than an intrinsic property of
the estimator.

The practical lesson is the same in both cases: a single-realization deviation from the
truth should always be weighed against the estimator sampling variance -- accessible via
repeated simulation when available, or via bootstrap/Monte-Carlo resampling on the real
data otherwise -- before being interpreted as evidence of a systematic bias. In
practice, this limitation is best addressed by increasing the number of sampled points
or independent realizations where possible, by widening $h$, or by a
bootstrap/Monte-Carlo calibration of the bias on surrogate data, in the spirit of
\pyzdcf own internal resampling.

Second, in contrast to \pyzdcf, which uses coarser, data-adaptive equal-population bins,
\nufftcf uses a dense, user-chosen lag grid evaluated pointwise. This gives much higher
lag resolution (Section~\ref{sec:comparison_ccf}) but, unlike \pyzdcf, does not itself
supply per-lag uncertainty estimates, which a user wanting formal error bars must still
obtain externally (e.g., by bootstrap).
%
\section{Summary and Conclusion}
\label{sec:conclusion}
%
We have presented \nufftcf, an open-source Python package that evaluates the Gaussian-
and rectangle-kernel ACF/CCF estimators of Section~\ref{sec:kernel_estimators} through
the Wiener--Khinchin theorem using the Non-Uniform FFT from \texttt{FINUFFT} library,
reducing their cost from $O(n^2)$ to $O(nK)$ for $K$ requested lags, i.e.\ linear in $n$
at fixed $K$, while remaining numerically consistent with their established real-space
definitions. An $O(n)$-per-lag two-pointer scan removes the $O(n^2)$ bottleneck in the
effective-pair-count normalization; it also becomes a major share of the cost of the
NUFFT estimators (dominant for the Gaussian kernel), whose spectral part is only
$O(n\log n)$. The real-space estimators, built on the same scan, share the $O(nK)$
scaling and remain the faster ones for a few lags, below a kernel-dependent threshold
$K^*$. Dedicated classic-FFT functions extend the same estimator definitions to
regularly-sampled data at a further reduced cost. All three estimator families share a
single calling convention, letting a user move between speed and an exact accuracy
reference without changing downstream code.

We validated \nufftcf against \pastas for ACF estimation and against \pyzdcf for CCF
estimation, on synthetic series with known analytical correlation structure, finding
agreement with both reference implementations within their respective statistical
precision (Section~\ref{sec:comparison}), and measured a linear-versus-quadratic scaling
advantage over \pastas ($\sim$30--90$\times$ at $n=8000$, and growing with $n$),
together with the dependence on the number of lags of the NUFFT-versus-real-space
trade-off, in direct timing benchmarks (Section~\ref{sec:benchmarks}). \nufftcf is
therefore positioned as a drop-in, order-of-magnitude faster alternative to existing
kernel-weighted ACF/CCF estimators for the long, dense irregularly-sampled time series
increasingly produced by continuous monitoring and high-cadence surveys, without
sacrificing the statistical grounding of the underlying estimator.

Beyond this controlled, cross-implementation validation, we also exercised \nufftcf on a
physically-motivated simulated use case (Section~\ref{sec:astro_example}): a stellar
photometric light curve combining a quasi-periodic rotation signal, correlated
``flicker'' noise, realistic ground-based sampling gaps, and heteroscedastic measurement
noise. \nufftcf, used alone, recovers both the injected rotation period and the noise
correlation time from the simulated observed series. Repeating this pipeline over
several independent realizations further showed that an apparent bias suggested by a
single realization was in fact a finite-sample selection effect rather than an intrinsic
property of the estimator (Section~\ref{sec:astro_mc}), reinforcing on an independent,
more realistic case the same conclusion reached from the synthetic peak-amplitude tests
of Section~\ref{sec:overestimation_peak_ampl}.

Future work may include extending \nufftcf uncertainty quantification beyond the
deterministic effective pair count $b$ following \pyzdcf own approach, exploring
additional weighting kernels within the same NUFFT reformulation. An other kind of
extension may be guided by the GPU version of \texttt{FINUFFT} \citep{cuFINUFFT2021} for
the very largest ($n\gtrsim10^6$) time series that may emerge from continuous
environmental and astronomical monitoring.

While the quantitative validation and benchmarks in this article rely on synthetic
series with known, controlled statistical properties---complemented by a single
physically motivated simulated light curve---the ultimate test of \nufftcf practical
utility lies in its application to both more realistic time series simulations, such as
those developed by \cite{Emmanoulopoulos2013}, and real-world data. Along this line, a
companion demonstration notebook\footnote{See
\texttt{nufftcf\_demo\_ccf\_Emmanoulopoulos.ipynb} notebook.}, whose results are not
detailed in this article, applies \nufftcf to such a simulated case, close to
reverberation mapping in high-energy astrophysics: an optical and a gamma-ray light
curve, sampled with different cadences, seasonal gaps and heteroscedastic measurement
errors, and offset by a known physical delay.

Finally, tests by the AGN reverberation-mapping community currently served by \pyzdcf,
and by hydrogeologists relying on \pastas for groundwater time series analysis, would
expose \nufftcf to observational irregularities --- gaps, heteroscedastic noise, and
non-stationary sampling --- that synthetic series may not fully capture, and would
constitute a more stringent test of \nufftcf than the synthetic validation performed in
this work.
%
\section{Reproducible Research}
\label{sec:data_availability}
The latest version of the \nufftcf package is available at
\url{https://github.com/jecampagne/nufftcf} under the MIT license, together with the
notebooks and benchmark scripts used to produce the comparisons and timings in this
article. The code is also available as a PyPI package at
\url{https://pypi.org/project/nufftcf/} to ease its installation. The documentation is
available at \url{https://jecampagne.github.io/nufftcf/}. For Coninuous Integration the
test suite is run via GitHub Actions on Linux, macOS (Apple Silicon/arm64 and
Intel/x86\_64), and Windows, across Python versions 3.11–3.14.

We have used \pastas 1.14.0, while concerning \pyzdcf we have made a fork of the
\url{https://github.com/LSST-sersag/pyzdcf} repository to allow the notebooks to be run
on the Google Colab plateform\footnote{\url{https://colab.research.google.com/}}.
%
\section*{Large Language Model Use Disclosure}
For the development of \nufftcf, we used the free version of Claude Desktop by Anthropic
(Sonnet 5\footnote{\url{https://www.anthropic.com/news/claude-sonnet-5}.}, Moyen) to
elaborate the Python project and set up the workflow for GitHub continuous integration
and systematic tests. We also used it to assist in writing the README and documenting
the notebooks and Python functions as well as their transcriptions into Latex
algorithms.

We also used Le Chat CNRS Enterprise, an AI-powered language model developed by Mistral
AI\footnote{\url{https://mistral.ai}} and tailored to the research needs of the National
Centre for Scientific Research (CNRS), to refine the clarity and fluency of the English
text.

All AI-generated suggestions were carefully reviewed and validated by the author.
%
\section*{Acknowledgments}
I would like to thank Julian Hamo for fruitful discussions, as well as the developers of
\texttt{FINUFFT}, the library upon which \nufftcf is based, \pastas, and \pyzdcf,
against which \nufftcf is validated.
%
\bibliographystyle{aasjournalv7}
\bibliography{refs_nufftcf}
%
\appendix
\section{NUFFT-based ACF/CCF algorithms}
\label{app:algorithms}
This appendix provides the complete pseudocode for the NUFFT-based estimators
implemented using the \texttt{FINUFFT} library, as outlined in
Section~\ref{sec:nufftcf_approach} (\texttt{nufft\_ccf.py} and \texttt{nufft\_acf.py}).
These estimators evaluate the Wiener--Khinchin identity of
Equation~\ref{eq:wienerkhinchin} via a type~1 NUFFT (from nonuniform samples to uniform
frequency grid) followed by a type~2 NUFFT (from uniform frequency grid to requested
lags). Both weighting kernels supported by \nufftcf (Gaussian and rectangle; see
Equations~\ref{eq:rectangle_kernel}--\ref{eq:Gaussian_kernel}) share the same
algorithmic structure. They differ only in the smoothing operator $\mathcal{S}_h$
(Gaussian filter of width $h$, or box filter of size $\mathrm{round}(2h)$) applied to
the raw NUFFT output, and in the corresponding pair-count function $b(\cdot)$ from
Appendix~\ref{app:pair-count}. Thus, each estimator is presented as a single,
kernel-parameterized algorithm to avoid redundancy.
\subsection{Cross and Autocorrelation functions}
Algorithm~\ref{alg:ccf-nufft} describes the general NUFFT-based CCF estimator. It
evaluates the Wiener--Khinchin identity by applying a type~1 NUFFT to each series,
computes the cross-spectrum $\hat X^\ast(f)\hat Y(f)$, evaluates it at the requested
lags using a type~2 NUFFT, smooths the result with the chosen kernel $\mathcal{S}_h$,
and normalizes it to a Pearson correlation in $[-1,1]$. This normalization uses the
matching effective pair count (Algorithm~\ref{alg:b-generic}) and the geometric mean of
the two series ACF values at lag $0$, $\sqrt{\mathrm{ACF}_x(0)\cdot\mathrm{ACF}_y(0)}$,
each obtained from Algorithm~\ref{alg:acf-scale0}.

The ACF (Algorithm~\ref{alg:acf-nufft}) is a special case of this estimator where $s=t$
and $y=x$. This simplifies the procedure in three ways. First, since only one series is
involved, the power spectrum $|\hat X(f)|^2$ is real and even, so only one type~1 NUFFT
call is required. The $\pm i$ sign convention, which must be carefully fixed for the CCF
cross-spectrum, is irrelevant here. Second, the effective pair count reduces to
Algorithm~\ref{alg:b-generic} evaluated with $s=t$. Third, normalization is simpler:
instead of computing the geometric mean of two scales
$\mathrm{ACF}_x(0)\cdot\mathrm{ACF}_y(0)$ separately (Algorithm~\ref{alg:acf-scale0}),
the ACF estimate is normalized directly by the smoothed value at lag $0$, obtained as
the first entry of the same evaluation array. The interior-position trick of
Algorithm~\ref{alg:acf-scale0} is unnecessary here because the ACF is even about
$\tau=0$.

\begin{algorithm*}
	\scriptsize
	\caption{Padded periodic domain and default frequency-grid size --- \textsc{PaddedDomain}}
	\label{alg:domain}
	\Require{$t_{\min}$, $\mathrm{span}$ (time span of the data; union of both series for the CCF); sorted evaluation lags $\tau^{\mathrm{srt}}$ (including lag $0$); $n_{\max}$ (number of samples of the longest series); optional user-supplied $N_1$}
	\Ensure{period $L$ of the NUFFT domain, frequency-grid size $N_1$, domain center $t_c$}
	$L \gets \mathrm{span} + 2\max_k|\tau^{\mathrm{srt}}_k|$ \tcp*{padded period: aliases of the requested lags fall outside the data}
	$t_c \gets t_{\min} + \mathrm{span}/2$ \tcp*{data are mapped onto a \emph{centered} arc of the circle}
	\If{$N_1$ \textnormal{not supplied}}{
		$N_1 \gets \mathrm{round}\big(32\,n_{\max}\,L/\mathrm{span}\big)$ \tcp*{Eq.~\ref{eq:N1_default}, Appendix~\ref{app:finufft-notes}}
	}
	\Return{$L, N_1, t_c$}
\end{algorithm*}

\begin{algorithm*}
	\scriptsize
	\caption{CCF via NUFFT + smoothing (Gaussian or rectangle)}
	\label{alg:ccf-nufft}
	\Require{times $t=(t^x_i)_{i=1}^{n_x}$, values $x=(x_i)$ (series 1,
		$t$ ascending); times $s=(t^y_j)_{j=1}^{n_y}$, values $y=(y_j)$
		(series 2, $s$ ascending); requested lags $\{\tau_1,\dots,\tau_K\}$ (any order);
		kernel filter $\mathcal{S}_h$ with bandwidth $h$ ($\sigma$ for the Gaussian kernel, half-width for
		the rectangle kernel); NUFFT tolerance $\varepsilon$; optional frequency-grid
		size $N_1$ (default given by Algorithm~\ref{alg:domain})}
	\Ensure{$c=(c_1,\dots,c_K)$ (CCF estimate, normalized to $[-1,1]$), $b=(b_1,\dots,b_K)$ (effective pair count), both in the order of the requested lags}
	$x_{\mathrm{std}} \gets (x-\bar x)/\mathrm{std}(x)$, \quad $y_{\mathrm{std}} \gets (y-\bar y)/\mathrm{std}(y)$ \tcp*{standardize both series}
	$\tau^{\mathrm{eval}} \gets (0,\tau_1,\dots,\tau_K)$, \quad $p \gets \mathrm{argsort}(\tau^{\mathrm{eval}})$, \quad $\tau^{\mathrm{srt}} \gets \tau^{\mathrm{eval}}[p]$ \tcp*{lag $0$ prepended for the normalization; sorted so that array and physical adjacency coincide for $\mathcal{S}_h$}
	$t_{\min} \gets \min(\min t,\min s)$, \quad $\mathrm{span} \gets \max(\max t,\max s) - t_{\min}$ \tcp*{common (union) time span}
	$(L,N_1,t_c) \gets \textsc{PaddedDomain}(t_{\min},\mathrm{span},\tau^{\mathrm{srt}},\max(n_x,n_y),N_1)$ \tcp*{Alg.~\ref{alg:domain}}
	$\tilde t \gets \pi + 2\pi\,\dfrac{t-t_c}{L}$, \quad
	$\tilde s \gets \pi + 2\pi\,\dfrac{s-t_c}{L}$, \quad
	$\tilde\tau \gets 2\pi\,\dfrac{\tau^{\mathrm{srt}}}{L}$\;
	$\hat F_1 \gets \textsc{nufft1d1}(\tilde t, x_{\mathrm{std}}, N_1,\varepsilon)$ \tcp*{type~1: time $\to$ frequency}
	$\hat F_2 \gets \textsc{nufft1d1}(\tilde s, y_{\mathrm{std}}, N_1,\varepsilon)$\;
	$M \gets \overline{\hat F_1}\odot \hat F_2$ \tcp*{cross-spectrum $\hat X^\ast(f)\hat Y(f)$, not Hermitian in general}
	$c_{\mathrm{raw}} \gets \Re\big[\textsc{nufft1d2}(\tilde\tau, M,\varepsilon)\big]$ \tcp*{type~2: frequency $\to$ lags}
	$c_{\mathrm{sm}} \gets \mathcal{S}_h(c_{\mathrm{raw}})$ \nllabel{ln:ccf-smooth} \tcp*{Gaussian or rectangle filtering along the sorted lag axis}
	$b_{\mathrm{cross}} \gets \textsc{PairCount}(s,t,\tau^{\mathrm{srt}},h,-1)$ \tcp*{Alg.~\ref{alg:b-generic} or~\ref{alg:b-rect-generic}, matching kernel, $t^y{=}s$, $t^x{=}t$}
	$c_{\mathrm{norm}} \gets c_{\mathrm{sm}} \oslash b_{\mathrm{cross}}$ \nllabel{ln:ccf-norm} \tcp*{elementwise division}
	$\mathrm{inv} \gets \mathrm{argsort}(p)$; \quad $c_{\mathrm{norm}} \gets c_{\mathrm{norm}}[\mathrm{inv}]$, \quad $b_{\mathrm{cross}} \gets b_{\mathrm{cross}}[\mathrm{inv}]$ \tcp*{back to the order $(0,\tau_1,\dots,\tau_K)$}
	$\mathrm{scale}_x \gets \textsc{AcfScaleAtLag0}(t,x_{\mathrm{std}},t_c,L,N_1,\varepsilon,h)$ \tcp*{Alg.~\ref{alg:acf-scale0}, same $(t_c,L,N_1)$ as the numerator}
	$\mathrm{scale}_y \gets \textsc{AcfScaleAtLag0}(s,y_{\mathrm{std}},t_c,L,N_1,\varepsilon,h)$\;
	$\mathrm{scale} \gets \sqrt{\mathrm{scale}_x \cdot \mathrm{scale}_y}$\;
	$c \gets c_{\mathrm{norm}}[2{:}] \,/\, \mathrm{scale}$ \tcp*{drop the lag-0 entry (indexing starts at 1)}
	$b \gets b_{\mathrm{cross}}[2{:}]$\;
	\Return{$c, b$}
\end{algorithm*}

\begin{algorithm*}
	\scriptsize
	\caption{ACF scale at lag $0$ (Pearson normalization) --- \textsc{AcfScaleAtLag0}}
	\label{alg:acf-scale0}
	\Require{$t$, $x_{\mathrm{std}}$, $t_c$, $L$, $N_1$ (the \emph{same} values as those used for the CCF numerator, see Alg.~\ref{alg:domain}), $\varepsilon$, bandwidth $h$}
	\Ensure{scalar estimate of $\mathrm{ACF}_x(0)$}
	$\tilde t \gets \pi + 2\pi\,(t-t_c)/L$\;
	$\hat F \gets \textsc{nufft1d1}(\tilde t, x_{\mathrm{std}}, N_1, \varepsilon)$, \quad $P \gets |\hat F|^2$\;
	$n_h \gets \max(\lceil 4h\rceil+2,\,5)$; \quad $\tau_{\mathrm{sym}} \gets (-n_h,\dots,-1,0,1,\dots,n_h)$ \nllabel{ln:scale0-nh} \tcp*{lag $0$ placed at an \emph{interior} array position}
	$\tilde\tau_{\mathrm{sym}} \gets 2\pi\,\tau_{\mathrm{sym}}/L$\;
	$c_{\mathrm{raw}} \gets \Re\big[\textsc{nufft1d2}(\tilde\tau_{\mathrm{sym}}, P,\varepsilon)\big]$\;
	$c_{\mathrm{sm}} \gets \mathcal{S}_h(c_{\mathrm{raw}})$ \tcp*{same smoothing as Alg.~\ref{alg:ccf-nufft}, applied symmetrically}
	$b_{\mathrm{sym}} \gets \textsc{PairCount}(t,t,\tau_{\mathrm{sym}},h,+1)$ \tcp*{Alg.~\ref{alg:b-generic} or~\ref{alg:b-rect-generic} with $s=t$}
	\Return{$c_{\mathrm{sm}}[n_h{+}1] \,/\, \max(b_{\mathrm{sym}}[n_h{+}1],10^{-16})$}
\end{algorithm*}

\begin{algorithm*}
	\scriptsize
	\caption{ACF via NUFFT + smoothing (Gaussian or rectangle)}
	\label{alg:acf-nufft}
	\Require{times $t$ ($n$ samples, ascending), values $x$; lags $\{\tau_1,\dots,\tau_K\}$ (any order); kernel filter $\mathcal{S}_h$ with bandwidth $h$; $\varepsilon$; optional $N_1$ (default given by Algorithm~\ref{alg:domain})}
	\Ensure{$c=(c_1,\dots,c_K)$, $b=(b_1,\dots,b_K)$, both in the order of the requested lags}
	$x_{\mathrm{std}} \gets (x-\bar x)/\mathrm{std}(x)$\;
	$\tau^{\mathrm{eval}} \gets (0,\tau_1,\dots,\tau_K)$, \quad $p \gets \mathrm{argsort}(\tau^{\mathrm{eval}})$, \quad $\tau^{\mathrm{srt}} \gets \tau^{\mathrm{eval}}[p]$\;
	$t_{\min} \gets \min t$, \quad $\mathrm{span} \gets \max t-\min t$\;
	$(L,N_1,t_c) \gets \textsc{PaddedDomain}(t_{\min},\mathrm{span},\tau^{\mathrm{srt}},n,N_1)$ \tcp*{Alg.~\ref{alg:domain}}
	$\tilde t \gets \pi + 2\pi\,\dfrac{t-t_c}{L}$, \quad
	$\tilde\tau \gets 2\pi\,\dfrac{\tau^{\mathrm{srt}}}{L}$\;
	$\hat F \gets \textsc{nufft1d1}(\tilde t, x_{\mathrm{std}}, N_1,\varepsilon)$\;
	$M \gets \hat F \odot \overline{\hat F} = |\hat F|^2$ \tcp*{real and even: the Hermitian case, no sign ambiguity}
	$c_{\mathrm{raw}} \gets \Re\big[\textsc{nufft1d2}(\tilde\tau, M,\varepsilon)\big]$\;
	$c_{\mathrm{sm}} \gets \mathcal{S}_h(c_{\mathrm{raw}})$ \nllabel{ln:acf-smooth}\;
	$b_{\mathrm{eval}} \gets \textsc{PairCount}(t,t,\tau^{\mathrm{srt}},h,+1)$ \tcp*{Alg.~\ref{alg:b-generic} or~\ref{alg:b-rect-generic}, $s=t$}
	$c_{\mathrm{eval}} \gets c_{\mathrm{sm}} \oslash b_{\mathrm{eval}}$\;
	$\mathrm{inv} \gets \mathrm{argsort}(p)$; \quad $c_{\mathrm{eval}} \gets c_{\mathrm{eval}}[\mathrm{inv}]$, \quad $b_{\mathrm{eval}} \gets b_{\mathrm{eval}}[\mathrm{inv}]$ \tcp*{back to the order $(0,\tau_1,\dots,\tau_K)$}
	$c \gets c_{\mathrm{eval}}[2{:}] \,/\, c_{\mathrm{eval}}[1]$ \tcp*{normalize by the lag-0 value}
	$b \gets b_{\mathrm{eval}}[2{:}]$\;
	\Return{$c,b$}
\end{algorithm*}
\subsection{Effective pair count algorithms}
\label{app:pair-count}
The pair-count functions $b(\cdot)$ (module \texttt{kernels.py}) are presented in
Algorithms~\ref{alg:b-generic}-\ref{alg:b-rect-generic}. Each algorithm computes the
\textit{effective} number of pairs contributing to each lag --- the normalization
denominator in Equation~\ref{eq:kernel_estimator}. They are implemented in
\texttt{numba} using a two-pointer scan: since both time arrays are sorted in ascending
order, the window $[\mathrm{lo},\mathrm{hi})$ associated with the pair-matching center
advances monotonically as the outer index increases. This reduces the computational cost
from $O(n_1 n_2)$ to $O(n_1+n_2)$ per lag, i.e.\ $O\big((n_1+n_2)K\big)$ for $K$ lags,
which are distributed over threads (\texttt{prange}).

The ACF and CCF pair-counts are \emph{not} separate algorithms but two instantiations of
the same two-pointer scan, differing only in the role of the time arrays and the sign
$\epsilon\in\{+1,-1\}$ in the window-center formula $c_j = t^{y}_j + \epsilon\,\tau_k$.
For the ACF ($t^{y}=t^{x}=t$, $\epsilon=+1$), the window center is $c_j = t_j+\tau_k$
with pair condition $t_i-t_j\approx\tau_k$, corresponding to
\texttt{compute\_b\_Gaussian} and \texttt{compute\_b\_rectangle}. For the CCF
($t^{y}=s$, $t^{x}=t$, $\epsilon=-1$), the window center is $c_j = s_j-\tau_k$ with pair
condition $s_j-t_i\approx\tau_k$, corresponding to \texttt{compute\_b\_Gaussian\_cross}
and \texttt{compute\_b\_rectangle\_cross}.

The sign flip encodes the asymmetry of the CCF (lag $+\tau$ means $y$ lags $x$ by
$\tau$; cf. the sign-convention note on the cross-spectrum). For the ACF, the sign
choice ($t_i-t_j\approx\tau_k$ rather than $t_j-t_i\approx\tau_k$) is an arbitrary but
fixed convention, harmless because $x$ is correlated with itself.
\begin{algorithm*}
	\scriptsize
	\caption{\textsc{PairCount} --- Gaussian (weighted) kernel, generic (ACF/CCF)}
	\label{alg:b-generic}
	\Require{$t^{y}=(t^{y}_j)_{j=1}^{n_y}$ sorted (outer array), $t^{x}=(t^{x}_i)_{i=1}^{n_x}$ sorted (window array), lags $\{\tau_k\}_{k=1}^{K}$, bandwidth $h$, sign $\epsilon\in\{+1,-1\}$}
	\Ensure{$b=(b_k)_{k=1}^K$}
	\ForPar{$k \gets 1$ \KwTo $K$}{
		$\mathrm{lo}\gets 0$, $\mathrm{hi}\gets 0$, $b_k \gets 0$\;
		\For{$j \gets 1$ \KwTo $n_y$}{
			\tcp{increasing $j$ $\Rightarrow$ increasing center $\Rightarrow$ lo/hi only advance}
			$\mathrm{center} \gets t^{y}_j + \epsilon\,\tau_k$\;
			\While{$\mathrm{lo}<n_x$ and $t^{x}_{\mathrm{lo}} < \mathrm{center}-6h$}{$\mathrm{lo}\gets \mathrm{lo}+1$\;}
			\While{$\mathrm{hi}<n_x$ and $t^{x}_{\mathrm{hi}} < \mathrm{center}+6h$}{$\mathrm{hi}\gets \mathrm{hi}+1$\;}
			\For{$i \gets \mathrm{lo}$ \KwTo $\mathrm{hi}-1$}{
				$b_k \gets b_k + \dfrac{1}{h\sqrt{2\pi}}\exp\!\Big(-\dfrac{(t^{x}_i-\mathrm{center})^2}{2h^2}\Big)$\;
			}
		}
		$b_k \gets \max(b_k, 10^{-16})$ \tcp*{avoids division by zero at sparsely-populated lags}
	}
	\Return{$b$}
	\smallskip
	\small $\bullet$ \textbf{ACF:} $t^{y}{=}t^{x}{=}t$, $n_y{=}n_x{=}n$, $\epsilon{=}{+1}$ $\Rightarrow$ \texttt{compute\_b\_Gaussian}.\quad
	$\bullet$ \textbf{CCF:} $t^{y}{=}s$, $t^{x}{=}t$, $\epsilon{=}{-1}$ $\Rightarrow$ \texttt{compute\_b\_Gaussian\_cross}.
\end{algorithm*}

\begin{algorithm*}
	\scriptsize
	\caption{\textsc{PairCount} --- rectangle (counting) kernel, generic (ACF/CCF)}
	\label{alg:b-rect-generic}
	\Require{$t^{y}=(t^{y}_j)_{j=1}^{n_y}$ sorted (outer array), $t^{x}=(t^{x}_i)_{i=1}^{n_x}$ sorted (window array), lags $\{\tau_k\}_{k=1}^K$, half-width $h$, sign $\epsilon\in\{+1,-1\}$}
	\Ensure{$b=(b_k)_{k=1}^K$}
	\ForPar{$k \gets 1$ \KwTo $K$}{
		$\mathrm{lo}\gets 0$, $\mathrm{hi}\gets 0$, $b_k \gets 0$\;
		\For{$j \gets 1$ \KwTo $n_y$}{
			$\mathrm{center} \gets t^{y}_j + \epsilon\,\tau_k$\;
			\While{$\mathrm{lo}<n_x$ and $t^{x}_{\mathrm{lo}} < \mathrm{center}-h$}{$\mathrm{lo}\gets \mathrm{lo}+1$\;}
			$\mathrm{hi} \gets \max(\mathrm{hi},\mathrm{lo})$\;
			\While{$\mathrm{hi}<n_x$ and $t^{x}_{\mathrm{hi}} \le \mathrm{center}+h$}{$\mathrm{hi}\gets \mathrm{hi}+1$\;}
			$b_k \gets b_k + (\mathrm{hi}-\mathrm{lo})$ \tcp*{direct count, no weighted sum: $O(1)$ per $j$}
		}
		$b_k \gets \max(b_k, 10^{-16})$\;
	}
	\Return{$b$}
	\smallskip
	\small $\bullet$ \textbf{ACF:} $t^{y}{=}t^{x}{=}t$, $\epsilon{=}{+1}$ $\Rightarrow$ \texttt{compute\_b\_rectangle}.\quad
	$\bullet$ \textbf{CCF:} $t^{y}{=}s$, $t^{x}{=}t$, $\epsilon{=}{-1}$ $\Rightarrow$ \texttt{compute\_b\_rectangle\_cross}.
\end{algorithm*}
\subsection{Implementation choices dictated by FINUFFT}
\label{app:finufft-notes}
Algorithms~\ref{alg:ccf-nufft}--\ref{alg:acf-nufft} rely on the \texttt{FINUFFT} library
for both nonuniform transforms (\texttt{nufft1d1}, type~1, and \texttt{nufft1d2},
type~2). Several implementation choices follow directly from this library constraints
and conventions. We detail them here, complementing the $O(n)$-per-lag \texttt{numba}
kernels of Appendix~\ref{app:pair-count}.
\paragraph{\bf Time domain and padded periodic circle.}
\texttt{FINUFFT} requires the nonuniform points to lie in one period of the transform,
so the sample times must be rescaled. Since the type~1 $\to$ type~2 round trip
implicitly assumes a periodic signal, we pad the period to
$L=\mathrm{span}+2\max_k|\tau_k|$ (Algorithm~\ref{alg:domain}) and map the data onto a
\emph{centered sub-arc} of the circle,
\begin{equation}
	\tilde t = \pi + 2\pi\,\frac{t-t_c}{L},
	\qquad t_c=t_{\min}+\frac{\mathrm{span}}{2},
	\label{eq:padded_map}
\end{equation}
the lags being rescaled as $\tilde\tau=2\pi\tau/L$. The circular alias of any requested
lag then falls outside the extent of the data, where it only multiplies by zero: this is
the NUFFT analogue of the zero-padding of the classic-FFT estimators, without any
fictitious zero-valued sample (which would bias the standardization). For
$\max_k|\tau_k|\to0$, Equation~\ref{eq:padded_map} reduces to the plain map onto the
whole circle. For the ACF, $\mathrm{span}$ is the extent of the series; for the CCF, $t$
and $s$ share the \emph{union} of their extents, $t_{\min}=\min(\min t,\min s)$ and
$\mathrm{span}=\max(\max t,\max s)-t_{\min}$, and the same map, so that the
cross-spectrum $\overline{\hat F_1}\odot\hat F_2$ is meaningful. The same $(t_c,L,N_1)$
must also be used for the lag-$0$ scale (Algorithm~\ref{alg:acf-scale0}); otherwise the
wrap-around bias would merely move from the numerator to the denominator.
\paragraph{\bf Sign convention and the non-Hermitian cross-spectrum.}
With \texttt{FINUFFT} default sign conventions, \texttt{isign=+1} for type-1
(Equation~\ref{eq:nufft_type1}) and \texttt{isign=-1} for type-2
(Equation~\ref{eq:nufft_type2}), this corresponds to the usual opposite-sign
Fourier-transform conventions. For the ACF, the power spectrum $|\hat X(f)|^2$ is real
and even, so the sign convention has no effect: $M=\hat F\odot\overline{\hat F}$. For
the CCF, the cross-spectrum $\hat X^\ast(f)\,\hat Y(f)$ \emph{is} Hermitian (since $x,y$
are real-valued), which guarantees that $c_{\mathrm{raw}}$ is real up to NUFFT roundoff.
However, it does not generally have exchange symmetry $x\leftrightarrow y$:
$$
S_{xy}(-f)=S_{yx}(f)\neq S_{xy}(f)\ \text{in general},
$$
mirroring the time-domain identity $C_{yx}(\tau)=C_{xy}(-\tau)$. Consequently, with the
convention $C_{xy}(\tau)=\int x(t)y(t+\tau)\,dt$, the choice
$M=\overline{\hat F_1}\odot\hat F_2$ places the correlation peak at the physical lag
$+\tau$ when signal 2 is delayed by $+\tau$ relative to signal 1, whereas the opposite
choice $\hat F_1\odot\overline{\hat F_2}$ effectively computes $S_{yx}(f)$ and therefore
places the peak at $-\tau$, i.e.\ a time-reversed (mirrored) result. This is a matter of
Fourier/CCF convention rather than numerical precision, but the convention must be
applied consistently for the sign of the estimated delay to have the intended physical
meaning.
\paragraph{\bf Frequency-grid size $N_1$.}
$N_1$ is the number of Fourier modes of the type~1 transform
(Algorithms~\ref{alg:ccf-nufft}--\ref{alg:acf-nufft}). On the padded period $L$ the
modes are spaced by $1/L$ and the highest resolved frequency is $f_{\max}=N_1/(2L)$. The
default is
\begin{equation}
	N_1=\mathrm{round}\!\left(32\,n_{\max}\,\frac{L}{\mathrm{span}}\right),
	\label{eq:N1_default}
\end{equation}
with $n_{\max}=\max(n_x,n_y)$ for the CCF and $n_{\max}=n$ for the ACF. The base factor
$32$ was validated empirically against the exact real-space estimator; the factor
$L/\mathrm{span}\geq1$ compensates the padding, so that
$f_{\max}=16\,n_{\max}/\mathrm{span}$ whatever the requested lags, and
$N_1\to32\,n_{\max}$ for $\max_k|\tau_k|\ll\mathrm{span}$. The extra cost is negligible
in that regime and at most a factor $\sim3$ for $\max_k|\tau_k|\simeq\mathrm{span}$
($L\simeq3\,\mathrm{span}$). A larger $N_1$ only marginally reduces the small residual
bias of the NUFFT estimators on strongly periodic, gappy signals (mainly for the
Gaussian kernel). The default is exposed as \texttt{default\_N1(n\_points, span, lags)};
an explicit \texttt{N1} bypasses both the base factor and the $L/\mathrm{span}$ scaling.
\paragraph{\bf Requested tolerance $\varepsilon$.}
The \texttt{eps} parameter sets the accuracy requested from \texttt{FINUFFT} internal
iterative scheme (gridding + FFT + deconvolution; Section~\ref{sec:nufft}). While
\texttt{FINUFFT} own library default is $10^{-6}$, our implementation uses a tighter
default of $10^{-9}$ (Algorithms~\ref{alg:ccf-nufft}--\ref{alg:acf-nufft}). Looser
values such as $10^{-6}$ speed up the computation but introduce NUFFT-specific numerical
noise.
\paragraph{\bf Lag-0 normalization and interior placement.}
The smoothing step (Gaussian or box filter) applied to the raw NUFFT output
(Algorithm~\ref{alg:ccf-nufft} step~8, Algorithm~\ref{alg:acf-nufft} step~6) introduces
an edge artifact: a point at the \emph{end} of the evaluated array is not smoothed
symmetrically, unlike an \emph{interior} point. The CCF peak of interest (at lag
$\tau_0$, interior to the requested \texttt{lags} range) \emph{is} an interior point.
Dividing this peak by an $\mathrm{ACF}(0)$ scale computed by placing lag $0$ at the
\emph{first} position of an array (hence smoothed asymmetrically) introduced a
systematic $2$--$3\,\%$ deficit in the estimated CCF peak.
Algorithm~\ref{alg:acf-scale0} (step~3) corrects this by evaluating $\mathrm{ACF}_x(0)$
on a small, \emph{symmetric} lag array centered on $0$ (indices
$-n_h,\dots,0,\dots,n_h$, with $n_h=\max(\lceil 4h\rceil+2,5)$, i.e.\ at least $4h$ of
margin on each side), so that the smoothing filter acts there exactly as it does at the
peak --- a concrete example of a numerical-pipeline artifact (not a property of the
underlying signal) that must be neutralized by how the evaluation array is constructed,
rather than corrected \textit{a posteriori}.
\paragraph{\bf Prior standardization.}
Both $x$ and $y$ are standardized (zero mean, unit variance) before any call to
\texttt{FINUFFT}. This is not dictated by \texttt{FINUFFT} itself, but ensures that the
NUFFT internal scale (which depends on input amplitude) is consistent between the two
compared series; combined with the final division by
$\sqrt{\mathrm{ACF}_x(0)\cdot\mathrm{ACF}_y(0)}$, it guarantees $c\in[-1,1]$ under the
Pearson convention, regardless of the original physical scale of either series.
\section{Real-space ACF/CCF algorithms}
\label{app:realspace-algos}
\begin{algorithm*}
		\scriptsize
		\caption{CorrSum --- Gaussian (weighted) kernel, generic (ACF/CCF)}
		\label{alg:c-generic}
		\Require{$t^{y}=(t^{y}_j)_{j=1}^{n_y}$ sorted (outer array), values $y=(y_j)$; $t^{x}=(t^{x}_i)_{i=1}^{n_x}$ sorted (window array), values $x=(x_i)$; lags $\{\tau_k\}_{k=1}^{K}$; bandwidth $h$; sign $\epsilon\in\{+1,-1\}$}
		\Ensure{$c=(c_k)_{k=1}^K$}
		\ForPar{$k \gets 1$ \KwTo $K$}{
			$\mathrm{lo}\gets 0$, $\mathrm{hi}\gets 0$, $c_k\gets 0$\;
			\For{$j \gets 1$ \KwTo $n_y$}{
				$\mathrm{center}\gets t^y_j + \epsilon\,\tau_k$\;
				\While{$\mathrm{lo}<n_x$ and $t^x_{\mathrm{lo}} < \mathrm{center}-6h$}{$\mathrm{lo}\gets \mathrm{lo}+1$\;}
				\While{$\mathrm{hi}<n_x$ and $t^x_{\mathrm{hi}} < \mathrm{center}+6h$}{$\mathrm{hi}\gets \mathrm{hi}+1$\;}
				$y_j' \gets y_j$\;
				\For{$i \gets \mathrm{lo}$ \KwTo $\mathrm{hi}-1$}{
					$c_k \gets c_k + \dfrac{1}{h\sqrt{2\pi}}\exp\!\left(-\dfrac{(t^x_i-\mathrm{center})^2}{2h^2}\right) x_i\, y_j'$\;
				}
			}
		}
		\Return{$c$}
		\smallskip
		\small $\bullet$ \textbf{ACF:} $t^y{=}t^x{=}t$, $y{=}x{=}x_{\mathrm{std}}$, $\epsilon{=}{+1}$ $\Rightarrow$ \texttt{compute\_c\_gaussian}.\quad
		$\bullet$ \textbf{CCF:} $t^y{=}s$, $y{=}y_{\mathrm{std}}$, $t^x{=}t$, $x{=}x_{\mathrm{std}}$, $\epsilon{=}{-1}$ $\Rightarrow$ \texttt{compute\_c\_gaussian\_cross}.
\end{algorithm*}

\begin{algorithm*}
		\scriptsize
		\caption{CorrSum --- rectangle kernel, generic (ACF/CCF), via prefix sums}
		\label{alg:c-rect-generic}
		\Require{$t^{y}=(t^{y}_j)_{j=1}^{n_y}$ sorted, values $y=(y_j)$; $t^{x}=(t^{x}_i)_{i=1}^{n_x}$ sorted, values $x=(x_i)$; lags $\{\tau_k\}_{k=1}^K$; half-width $h$; sign $\epsilon\in\{+1,-1\}$}
		\Ensure{$c=(c_k)_{k=1}^K$}
		$\mathrm{cumsum}_x[0]\gets 0$; \quad $\mathrm{cumsum}_x[i]\gets \mathrm{cumsum}_x[i-1]+x_i$ for $i=1,\dots,n_x$ \tcp*{precomputed once}
		\ForPar{$k \gets 1$ \KwTo $K$}{
			$\mathrm{lo}\gets 0$, $\mathrm{hi}\gets 0$, $c_k\gets 0$\;
			\For{$j \gets 1$ \KwTo $n_y$}{
				$\mathrm{center}\gets t^y_j + \epsilon\,\tau_k$\;
				\While{$\mathrm{lo}<n_x$ and $t^x_{\mathrm{lo}} < \mathrm{center}-h$}{$\mathrm{lo}\gets \mathrm{lo}+1$\;}
				$\mathrm{hi}\gets\max(\mathrm{hi},\mathrm{lo})$\;
				\While{$\mathrm{hi}<n_x$ and $t^x_{\mathrm{hi}} \le \mathrm{center}+h$}{$\mathrm{hi}\gets \mathrm{hi}+1$\;}
				$c_k \gets c_k + y_j\big(\mathrm{cumsum}_x[\mathrm{hi}]-\mathrm{cumsum}_x[\mathrm{lo}]\big)$ \tcp*{$O(1)$ per $j$, uniform kernel weight}
			}
		}
		\Return{$c$}
		\smallskip
		\small $\bullet$ \textbf{ACF:} $t^y{=}t^x{=}t$, $y{=}x{=}x_{\mathrm{std}}$, $\epsilon{=}{+1}$ $\Rightarrow$ \texttt{compute\_c\_rectangle}.\quad
		$\bullet$ \textbf{CCF:} $t^y{=}s$, $y{=}y_{\mathrm{std}}$, $t^x{=}t$, $x{=}x_{\mathrm{std}}$, $\epsilon{=}{-1}$ $\Rightarrow$ \texttt{compute\_c\_rectangle\_cross}.
\end{algorithm*}

\begin{algorithm*}
		\scriptsize
		\caption{ACF via real-space kernel-weighted summation (Gaussian or rectangle)}
		\label{alg:acf-realspace}
		\Require{times $t$ ($n$ samples, ascending), values $x$; lags $\{\tau_1,\dots,\tau_K\}$; kernel bandwidth $h$ ($\sigma$ for Gaussian, half-width for rectangle)}
		\Ensure{$c=(c_1,\dots,c_K)$, $b=(b_1,\dots,b_K)$}
		$x_{\mathrm{std}} \gets (x-\bar x)/\mathrm{std}(x)$\;
		$c_{\mathrm{raw}} \gets \textsc{CorrSum}(t,x_{\mathrm{std}},t,x_{\mathrm{std}},(\tau_1,\dots,\tau_K),h,+1)$ \tcp*{Alg.~\ref{alg:c-generic} or~\ref{alg:c-rect-generic}}
		$b \gets \textsc{PairCount}(t,t,(\tau_1,\dots,\tau_K),h,+1)$ \tcp*{Alg.~\ref{alg:b-generic} or~\ref{alg:b-rect-generic}, matching kernel}
		$c \gets c_{\mathrm{raw}} \oslash b$\;
		\Return{$c, b$}
\end{algorithm*}

\begin{algorithm*}
		\scriptsize
		\caption{CCF via real-space kernel-weighted summation (Gaussian or rectangle)}
		\label{alg:ccf-realspace}
		\Require{times $t$, values $x$ (series 1, $t$ ascending); times $s$, values $y$ (series 2, $s$ ascending); lags $\{\tau_1,\dots,\tau_K\}$; kernel bandwidth $h$}
		\Ensure{$c=(c_1,\dots,c_K)$ (CCF estimate, normalized to $[-1,1]$), $b=(b_1,\dots,b_K)$ (effective pair count)}
		$x_{\mathrm{std}} \gets (x-\bar x)/\mathrm{std}(x)$, \quad $y_{\mathrm{std}} \gets (y-\bar y)/\mathrm{std}(y)$\;
		$\tau^{\mathrm{eval}} \gets (0,\tau_1,\dots,\tau_K)$ \tcp*{lag $0$ prepended; no sort needed (no smoothing-filter adjacency to preserve)}
		$c_{\mathrm{cross}} \gets \textsc{CorrSum}(s,y_{\mathrm{std}},t,x_{\mathrm{std}},\tau^{\mathrm{eval}},h,-1)$ \tcp*{Alg.~\ref{alg:c-generic} or~\ref{alg:c-rect-generic}}
		$b_{\mathrm{cross}} \gets \textsc{PairCount}(s,t,\tau^{\mathrm{eval}},h,-1)$ \tcp*{Alg.~\ref{alg:b-generic} or~\ref{alg:b-rect-generic}}
		$c_{\mathrm{norm}} \gets c_{\mathrm{cross}} \oslash b_{\mathrm{cross}}$\;
		$\mathrm{scale}_x \gets \textsc{CorrSum}(t,x_{\mathrm{std}},t,x_{\mathrm{std}},(0),h,+1) \,/\, \textsc{PairCount}(t,t,(0),h,+1)$ \tcp*{ACF$_x(0)$, direct evaluation}
		$\mathrm{scale}_y \gets \textsc{CorrSum}(s,y_{\mathrm{std}},s,y_{\mathrm{std}},(0),h,+1) \,/\, \textsc{PairCount}(s,s,(0),h,+1)$\;
		$\mathrm{scale} \gets \sqrt{\mathrm{scale}_x \cdot \mathrm{scale}_y}$\;
		$c \gets c_{\mathrm{norm}}[2{:}] \,/\, \mathrm{scale}$ \tcp*{drop the lag-0 entry}
		$b \gets b_{\mathrm{cross}}[2{:}]$\;
		\Return{$c, b$}
\end{algorithm*}

Algorithms~\ref{alg:c-generic}--\ref{alg:ccf-realspace} give the pseudocode for the
real-space estimators (\texttt{realspace\_acf.py}, \texttt{realspace\_ccf.py}), the
direct counterpart of the NUFFT-based Algorithms~\ref{alg:ccf-nufft}
and~\ref{alg:acf-nufft}. Because they evaluate Equation~\ref{eq:kernel_estimator} by a
direct pairwise sum rather than via the Wiener--Khinchin identity, none of the
FINUFFT-specific machinery of Appendix~\ref{app:finufft-notes} is needed: no padded
periodic domain, no sign convention, no frequency-grid size $N_1$, and no interior-lag
correction for the lag-0 scale (Algorithm~\ref{alg:acf-scale0}). This is precisely what
makes the real-space path an artifact-free reference.

The correlation numerator is computed by \textsc{CorrSum}
(Algorithms~\ref{alg:c-generic}--\ref{alg:c-rect-generic}), the direct analogue of the
pair-count denominator \textsc{PairCount} (Algorithms~\ref{alg:b-generic}
and~\ref{alg:b-rect-generic}): both run the same two-pointer scan over the sorted time
arrays, \textsc{CorrSum} accumulating the weighted product $x_i\,y_j$ where
\textsc{PairCount} accumulates the kernel weight alone.
Algorithms~\ref{alg:acf-realspace} and~\ref{alg:ccf-realspace} then assemble $c$ and $b$
from one or two calls to each, without the lag-sorting step required by the NUFFT path's
smoothing filter $\mathcal{S}_h$ (Algorithm~\ref{alg:ccf-nufft},
line~\ref{ln:ccf-smooth}), since no adjacency needs to be preserved along the lag axis
here.
\end{document}